\documentclass[sigconf]{acmart}

\AtBeginDocument{%
  }

\copyrightyear{2023}
\acmYear{2023}
\setcopyright{acmlicensed}\acmConference[CHI '23]{Proceedings of the 2023
CHI Conference on Human Factors in Computing Systems}{April 23--28,
2023}{Hamburg, Germany}
\acmBooktitle{Proceedings of the 2023 CHI Conference on Human Factors in
Computing Systems (CHI '23), April 23--28, 2023, Hamburg, Germany}
\acmPrice{15.00}
\acmDOI{10.1145/3544548.3581173}
\acmISBN{978-1-4503-9421-5/23/04}

\usepackage{multirow}
\usepackage{makecell}

\begin{document}

\title[Making Sense of Roles in Web3 Community from Airdrop Perspective]{Altruistic and Profit-oriented: Making Sense of Roles in \\ Web3 Community from Airdrop Perspective}

\author{Sizheng Fan}
\affiliation{%
  \institution{The Chinese University of Hong Kong, Shenzhen}
  \city{Shenzhen}
  \country{China}}
\email{sizhengfan@link.cuhk.edu.cn}

\author{Tian Min}
\affiliation{%
  \institution{The Chinese University of Hong Kong, Shenzhen}
  \city{Shenzhen}
  \country{China}}
\email{tianmin@link.cuhk.edu.cn}

\author{Xiao Wu}
\affiliation{%
  \institution{White Matrix Inc.}
  \city{Nanjing}
  \country{China}}
\email{wuxiao@whitematrix.io}

\author{Wei Cai}
\authornote{Wei Cai is the corresponding author.}
\affiliation{%
  \institution{The Chinese University of Hong Kong, Shenzhen}
  \city{Shenzhen}
  \country{China}}
\email{caiwei@cuhk.edu.cn}


\begin{abstract}

Regardless of which community, incentivizing users is a necessity for well-sustainable operations. In the blockchain-backed Web3 communities, known for their transparency and security, airdrop serves as a widespread incentive mechanism for allocating capital and power. However, it remains a controversy on how to justify airdrop to incentive and empower the decentralized governance. In this paper, we use ParaSwap as an example to propose a role taxonomy methodology through a data-driven study to understand the characteristic of community members and the effectiveness of airdrop. We find that users receive more rewards tend to take positive actions towards the community. We summarize several arbitrage patterns and confirm the current detection is not sufficient in screening out airdrop hunters. In conjunction with the results, we discuss from the aspects of interaction, financialization, and system design to conclude the challenges and possible research directions for decentralized communities.
\end{abstract}

\ccsdesc[500]{Human-centered computing~Empirical studies in collaborative and social computing}

\keywords{Decentralized community, Airdrop, Network analysis, Unsupervised learning}

\maketitle

\section{Introduction}
\label{sec:Introduction}
Since the early days, many research efforts have been on evaluating a community, trying to model its members and quantify metrics such as reciprocity and sustainability. As technology has evolved, communities have come to refer to more than just real people, but can also be used to represent groups of users on the Internet that share specific characteristics. With the development of blockchain technology and the popularization of decentralized applications, the emergence of the so-called \emph{Web3} has brought the communities built on this particular background into our research horizon. The anonymity of blockchain prevents us from conducting user interviews and questionnaire research, but its unique transparent nature allows us to infer user behavior through publicly available datasets.

Compared to centralized communities, decentralized communities of Web3 introduce more challenges, including ways to rationalize the allocation of power and resources or ensure that individuals do not dictate the will of the community. The disappearing central governance reforms itself in various mechanisms of other composite appearances, such as democratic voting on community development using the governance token. A real-world example is Juno Network, published \emph{Juno Proposal 16} on March 11$^{th}$, 2022, calling the community to whether the wallet of a whale\footnote{Means an address that holds a large amount of capital.} should have a large chunk of its token removed, like court enforcement of property. The motivation behind this was when Juno Network rewarded its early supporters with governance token, known as \emph{an airdrop}\footnote{An airdrop is an unsolicited distribution of cryptocurrency or Non Fungible Tokens (NFTs), usually without conditions, to multiple wallet addresses.}, the whale obtained 2.5 million tokens through multiple addresses, which accounted for 9.6\% of the total amount and could bring a huge potential security risk to the community. The proposal is highly controversial because it marks the first strong-arm tactics of on-chain governance. It is high time for decentralized communities to discuss and rethink the essence of their own creation. On the one hand, it raises the community's concern about the ownership of on-chain assets. On the other hand, the potential harm and future development of the airdrop have been widely discussed.

Unlike a community in the traditional context, a decentralized community is much looser in scope. Joining and exiting a decentralized community often simply means whether you are a stakeholder of a certain type of token: holding assets issued by a specific protocol implies corresponding rights and obligations. The protocols, substantially, are the large number of automatically executed programs not controlled by individuals or organizations, also called \emph{smart contracts}, tasked with the responsibility of regulating the community. The regulation or resolution for the decentralized community is valid only if most of its members support the decision through fungible tokens (FTs) or NFTs voting. These \emph{governance tokens} grant their holders voting rights like a shareholder meeting of listed companies. Members and potential participants are usually motivated to contribute to the community, especially in the early stages. For example, some decentralized applications (DApps), such as Uniswap, dYdX, etc., issue airdrops to early supporters for their contributions. Some NFT projects offer whitelist quota\footnote{A whitelist in the NFT world is a list of people who can get early and guaranteed access to mint during a specific date.} for promotion services. The simple logic behind this is the contribution for the privilege, which indicate that users who contribute more are more likely to be competent to the governance and should be given more voting power.

However, due to the financial nature inherent in blockchain technology the blockchain, as of now, all digital assets on the blockchain network can be exchanged for real-world currency by various means, which implies that from a gaming standpoint, decentralized communities in Web3 are facing a more significant challenge in terms of altruism than traditional Internet communities with open source spirit. Among the early supporters eligible for airdrop, most of them are motivated by financial profit or preferential access to tokens with governance rights in order to obtain quick cash. Many of them exploit the anonymity of blockchain, registering multiple accounts and interacting with the DApp to maximize their profits, which can not be taken as a positive factor for the long-term development, because once they acquire the tokens, exploiters tend to sell the tokens as soon as possible, consequently making a drop of the asset price. These behaviors undermine the rights and the interests of those enthusiasts who would like to hold the tokens in the long run and proactively engage in the community.

To address these issues, many DApps have begun to distribute governance tokens by designing stringent filtering mechanisms where a typical example is ParaSwap\footnote{ParaSwap is a decentralized middleware aggregator on the Ethereum blockchain that offers the best prices across various DEXs.}. It is particularly strict in selecting eligible airdrop addresses: in the past two years, around 1.3 million addresses have interacted with this protocol, but only 19,999 of them were rewarded with PSP, ParaSwap's governance token. The rigorous filtering approach caused widespread concern across the Web3 community. Some early supporters criticized the project for being screened out while they are actual users who made a practical contribution. The distribution method is significant because of its dual implication in terms of governance power and monetary, and thus triggered intense debate on the effectiveness of airdrops.

In this paper, we take ParaSwap as a representative example, trying to evaluate the Web3 community and the effectiveness of allocation principles through the analysis of eligible users' behavior and token transaction network. We collect PSP transactions of 20,148 addresses between November 15$^{th}$, 2021 and April 13$^{th}$, 2022, based on which we analyze their behavior before and after the airdrop event and the allocation method from the following two aspects. From the macro aspect, we investigate the temporal development of the network properties, based on which we also find a series of the unique structure of the interaction between token holders through component analysis by comparing the token network with the external transaction network. From the micro aspect of individual addresses, we perform unsupervised clustering on eligible airdrop addresses based on their transaction patterns. We then compare the distribution difference in terms of the amount of capital and token holding duration among different clusters. Our main contributions can be summarised as follows:

\begin{itemize}
    \item \emph{Token Network Analysis.} 
    We employ global network properties, such as reciprocity, degree assortativity coefficient, and attracting components, to detect significant changes and anomalies over PSP token networks. We decompose the network into components and discuss it in-depth with respect to potential arbitrage behavior.
    
    \item \emph{Unsupervised Clustering}. 
    We utilize an unsupervised hierarchical clustering method to capture the behaviors of eligible airdrop addresses after receiving rewards and divide them into interpretable categories. Based on the cluster results, we reveal that the threshold differential allocation can perform a better allocation of governance power to long-term contributors.
    
    \item \emph{Evaluation on Web3 Community}.
    As a prerequisite for rational governance and creating a healthy online community, the suitable incentive mechanism and power allocation play an irreplaceable role. We point out that huge problems persist behind the idealism of equality and democracy promoted by concepts such as the \emph{Metaverse} \cite{metaverse_2}. Combined with our data findings, we discuss the pros and cons of current approaches and suggest that social experiments in blockchain and Web3 as vehicles may be instructive for real-world social research. Finally, we proposed the challenges of the community engagement and governance as future directions and potential improvements for the Web3 Community.
\end{itemize}

\section{Background}\label{sec:background}
\subsection{Ethereum and Smart Contract}
\label{sec:ethereum_and_Web 3.0}
Ethereum \cite{article:EthereumYellowPaper, 8345547} has a long history of development and was designed to address several limitations and challenges of the Bitcoin \cite{nakamoto2008peer}. It provides the developers with a tightly integrated end-to-end system for building software on a hitherto unexplored compute paradigm in the mainstream: a trustful object messaging compute framework with smart contracts, which are scripts that run synchronously on multiple nodes of a distributed ledger without the need of an external trusted authority \cite{8847638, swan2015blockchain}. At the inception of Ethereum, it followed the consensus mechanism\footnote{Consensus mechanism refer to methodologies used to achieve consistent, trust, and security across a decentralized computer network.} of Bitcoin, called Proof-of-Work (PoW), which requires network members to solve an arbitrary mathematical puzzle. As PoW has the problem of wasting a lot of computing power with low performance and efficiency, blockchain platforms have started to adopt more efficient models, including Proof-of-Stake (PoS), Delegated Proof-of-Stake (DPoS), and Proof-of-Authority (PoA) \cite{han2022can} as the technology evolves. Ethereum upgraded from the PoW to PoS on September 15, 2022, so-called \emph{The Merge}\footnote{https://ethereum.org/en/upgrades/merge/}, which has dramatically improved its transaction per second (TPS) and further facilitated the development of DApps.


There are three types of addresses on Ethereum: 1) \emph{Externally Owned Address}, so-called user address, operated by human users with private keys; 2) \emph{Contract Address}, namely the smart contracts, governed by the internal contract code, acting as an autonomous agent; 3) \emph{Prescribed Address} is a kind of special addresses with various functionalities that similar to the prescribed strings in a programming language.

Judging by the identity of the party who initiates the transaction, we can divide the transaction types into two categories: \emph{external} and \emph{internal}. External transactions are initiated by user addresses, and they can be direct transactions between user addresses or function calls to smart contracts. Internal transactions are initiated by smart contracts that can transmit between themselves or can send tokens to users, as in the case of airdrop contracts.


\subsection{Airdrop in Web3}
\label{sec:airdrop_in_web3.0}

Prior to delving into the topic of governance in the realm of Web3, it is imperative to have a basic understanding of the formation process of decentralized communities. This will facilitate a clearer appreciation of the crucial part played by airdrops in the enhancement of community empowerment and decentralization progression.

The empowerment of decentralized communities in Web3 is a gradual and long-term process. Similar to public companies, DApp start-up teams follow a defined process to publish their white paper, which outlines the total number of tokens and their allocation, as well as the timeline for distribution. Although the regulations may vary from DApp to DApp, the format tends to be standardized. The majority of DApps complete the distribution process within a few years, excluding the community vault and tokens retained by the start-up team. The method of distribution is crucial to the success of Web3 communities, as the absence of this process would render the community an incomplete and unfulfilled vision. To ensure the success and sustainability of decentralized communities, it is essential to have a well-defined and executed plan for token distribution and community empowerment.

Airdrops, as a mainstream way of distributing tokens, are initiated by crypto projects to offer their native tokens to current or potential users in building a decentralized community. They serve multiple purposes, being akin to virtual coupons that can be deposited directly into a wallet. These coupons serve as a way to attract new customers or reward existing ones. The tokens that are a part of an airdrop are usually compliant with Ethereum Request for Comments (ERC)\footnote{ERC is a document to write smart contracts on Ethereum. They describe rules that Ethereum-based tokens must comply with.} standards and can function as a form of currency or grant governance rights in the form of governance tokens. The distribution of airdrops is determined by the level of participation in specific events hosted by the blockchain platform, DApps, or NFT purchases. In conclusion, airdrops play a vital role in fostering the growth and involvement of decentralized communities and are, therefore, a crucial tool for any crypto project seeking to establish a thriving decentralized community.



\subsection{ParaSwap's Allocation Method}

\begin{figure}[hbp]
    \centering
    \includegraphics[width=1\columnwidth]{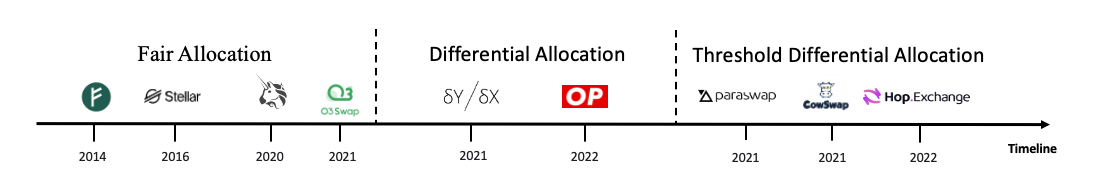}
    \caption{Development of Airdrop}
    \label{fig:airdrop_timeline}
    \Description{It is a timeline describing the development of Airdrop, divided into three phases: Fair Allocation, Differential Allocation and Threshold Differentail Allocation. each development phase is marked with the logo of the corresponding DApp at the bottom.}
\end{figure}

As more and more DApps are using airdrops to reward early adopters with increasingly large amounts of money, a breed of people known as \emph{"Airdrop Hunters"}, who create multiple wallet addresses and interact with the protocol to obtain multiple airdrops \cite{8325269}. Their actions will hinder the decentralization of the protocol. To cope with this situation, DApps are constantly optimizing and adjusting their allocation policies. As shown in Figure \ref{fig:airdrop_timeline}, from fair allocation led by Uniswap in the very beginning, to differential allocation trend formed by dYdX\footnote{https://dydx.exchange}, to threshold differential allocation represented by ParaSwap, the difficulty of getting airdrop is increasing. This is an important reason why we choose ParaSwap as the object of our study, because it is not only the representative of airdrop method at this stage, but its stringent filter is also the trend of future development.

To identify the hunters, ParaSwap first filtered the addresses with at least 50 transactions on the respective networks or have a minimum native token balance\footnote{ParaSwap set the minimum token balance to 0.028 ETH for Ethereum, 0.25 BNB for BSC, 20 Matic for Polygon, and 0.9 AVAX for Avalanche C-Chain.}. Different from existing rules, ParaSwap set a threshold of having at least six times of interactions in the last six months. Besides, ParaSwap leveraged the common patterns between hunters since they usually make external transactions between their accounts, which could filter addresses which were a part of a clique with size \textgreater 5. User will be graded with their level of activity to determine the reward tier\footnote{https://medium.com/paraswap/whats-an-active-user-clarifying-psp-token-distribution-filtering-logic-81df6096d410}.
 



\section{Related Work}\label{sec:related_work}

\subsection{Governance in Online Communities}
Millions of communities gather in online spaces such as Twitter, Reddit, Discord, and so on, which fill an essential part of our daily lives and continue to shape an interconnected social scene differently than ever before. However, behind the prosperity of online communities, issues such as disinformation, sexual harassment, inflammatory and hate speech remain. Therefore, regulations are urgently needed and play an irreplaceable role in community governance. Today, community governance is mainly in the form of a centralized model consisting of roles and permissions, where groups such as administrators and moderators have broad privileges over ordinary users. The regulators can be a specific department of service provider, user autonomy \cite{web2ModerationVisibility}, or hired personnel for content review \cite{cai2019categorizing}, to perform intervention or censorship such as deplatforming \cite{web2Deplatforming}, quarantining \cite{web2Quarantined} with the help of artificial intelligence and algorithms \cite{articleModeratorEngagement, web2HumanMachineCollaboration}. Originating from Web 2.0 developed nearly twenty years ago, this centralized governance model is now used in software for almost all major online platforms. Researchers have conducted an extensive and in-depth study on the governance aspects of online communities. For example, Manoel \emph{et al.} \cite{web2PlatformMigration} studied the level of activity, and ideological changes, when toxic online communities on mainstream platforms faced moderation measures such as bans and migrated to other platforms. They showed increases in signals associated with toxicity and radicalization, which justifies concerns that the reduction in activity may come at the expense of a more toxic and radical community. Faced with an endless chase between moderator and violator, researchers are exploring the potential of blockchain technology to facilitate community governance and democratize decision-making.

By autonomously executing smart contracts, DApps can provide users with a variety of decentralized services, including crowdfunding, decentralized exchanges (DEXs), social networking, decentralized autonomous organizations (DAOs), and blockchain games. They expand the possibility of governance and encode certain values that make top-down, authoritarian and, punitive governance easier to implement. DApps try to achieve a complete decentralization and direct democracy through issuing governance token, that is, to establish a universal voting system where the direction of the project's development is determined by the whole community. Proposals for voting can be community cooperation, using the community vault, or changing the rule of a digital game. The governance under blockchain is challenging, and it is difficult to steer a decentralized community and promote its development without sacrificing decentralization \cite{OverviewOfDAO}. Techno-determinism seems more favored at the expense of the complexity of the social organization. This has partly contributed to the current chaotic public opinion on blockchain. Proponents of centralization also argue that a traditional centralized authority is necessary for the current highly financialized blockchain to emphasize the role of the state \cite{IsTheStateStillNecessary}.

Although in the very recent past, blockchain-based media platforms (LBRY, CreativeChain, MediaChain, Blockphase) emerged to provide services such as copyright protection \cite{elsden2018making}, the emphasis of decentralized community governance is much different compared to the mainstream Web 2.0 community nowadays. Due to the disparity in user volume and platform traffic, it is difficult for us to identify social platforms similar to Twitter and Facebook from Web3 and to draw a side-by-side comparison of their mechanism in content regulation. However, this does not prevent us from exploring the characteristics of user behavior in existing decentralized communities to pave the way for possible future emergence.

\subsection{Study on Decentralized Communities}
\label{sec:study_on_decentralized_communities}
Since the inception of blockchain technology, developers and researchers have never stopped trying to define or paint an ideal picture of a decentralized community, also known as DAOs, in terms of technology and social science benchmarks, but there does not have a consensus. However, a comprehensive definition made by Luis Cuende is that “a DAO is an internet-native entity with no central management which is regulated by a set of automatically enforceable rules on a public blockchain, and whose goal is to take a life of its own and incentives people to achieve a shared common mission” \cite{OverviewOfDAO}.

Lustig \emph{et al.} \cite{IntersectingImaginaries} studied the sociotechnical imaginaries about how people envision the possible and desirable future of decentralized autonomous systems and identified three framings of autonomous technology as physical objects, as mathematical rules and as artificial managers, which can inform the design and governance of this novel concept. Faqir \emph{et al.} \cite{OverviewOfDAO}, from a more specific and practical perspective, review the collective governance DAO providing platforms and their key features. At the same time, from the aspect of system construction, DAO-Analyzer-like evaluation systems \cite{DAO-Analyzer} that can visualize the activity and participation of DAO are also being actively developed, thanks to the open-source nature of blockchain data. At the same time, other critical and skeptical voices are helping to correct the direction of the decentralized community. Nabben \emph{et al.} \cite{web3DAOPanopticonVulnerability} suggested that reputation in blockchain systems could become the new algorithmic authoritarianism if misused for social control and would conversely weaken the human factor in autonomy by employing ethnographic methods. Meijer \emph{et al.} \cite{InstitutionalBlockchainGovernance} provide a conceptualization of the consequences of blockchain systems that can imply the way of developing regulatory arrangements and strategies that strike a balance between power and possibilities that blockchain applications offer to mitigate negative effects.

While other previous research focuses on the data-driven analysis on transaction records and token flow \cite{10.1145/3555858.3555883, fan2022towards, min2022portrait}, but only few of them concentrate on the specific topic of decentralized communities. Friedhelm Victor \emph{et al.} \cite{victor2019measuring} presented a descriptive measurement study to reveal the quantitative characters of the token network by providing an overview of more than 64,000 ERC-20 token networks and pointed out that even though the token network on Ethereum has followed a power law in its degree distribution, lots of individual token networks are frequently dominated by spoke pattern or a single hub. Similarly, Chen \emph{et al.} \cite{chen2020traveling} analyzed the whole token ecosystem on Ethereum with emphasis on the token creators and stakeholders. The result suggested that token holding and token creation are profoundly unbalanced. These studies can be based to some extent on our expectations about the distribution of governance tokens, which, as mentioned in Section \ref{sec:Introduction}, are the materialization of the rights of community members. The imbalance in the distribution of power indicates a so-called decentralized community is substantially controlled by some minority groups of authority, which further implies deficiencies in the process of community empowerment. Furthermore, airdrop, the most dominant means of power allocation, has not been widely evaluated yet.

We consider our study a tentative attempt to derive the rationality of the airdrop model backward from the results regarding users' behavior after acquiring them. And more importantly, we try to set up a research pipeline of role identification to understand and conclude the users' composition and behavioral patterns in a decentralized community. We hope this work could provide referable information and data-driven evidence to support the further design and research on decentralized governance models.

\section{Dataset Overview}
\label{sec:datasets_and_experimental_setup}

As we mentioned in Section \ref{sec:Introduction}, the Web3 community has very loose boundaries, i.e., as long as having a corresponding token, anyone can be accepted as a member. For the sake of brevity and precision, in the rest of this paper, we will use \emph{Member} to refer to all addresses holding PSP tokens in general and \emph{Initial Member} to represent the 13,830 addresses that initially qualified for and claimed the airdrop.

\begin{figure}[htbp]
	\centering
    \includegraphics[width=\columnwidth]{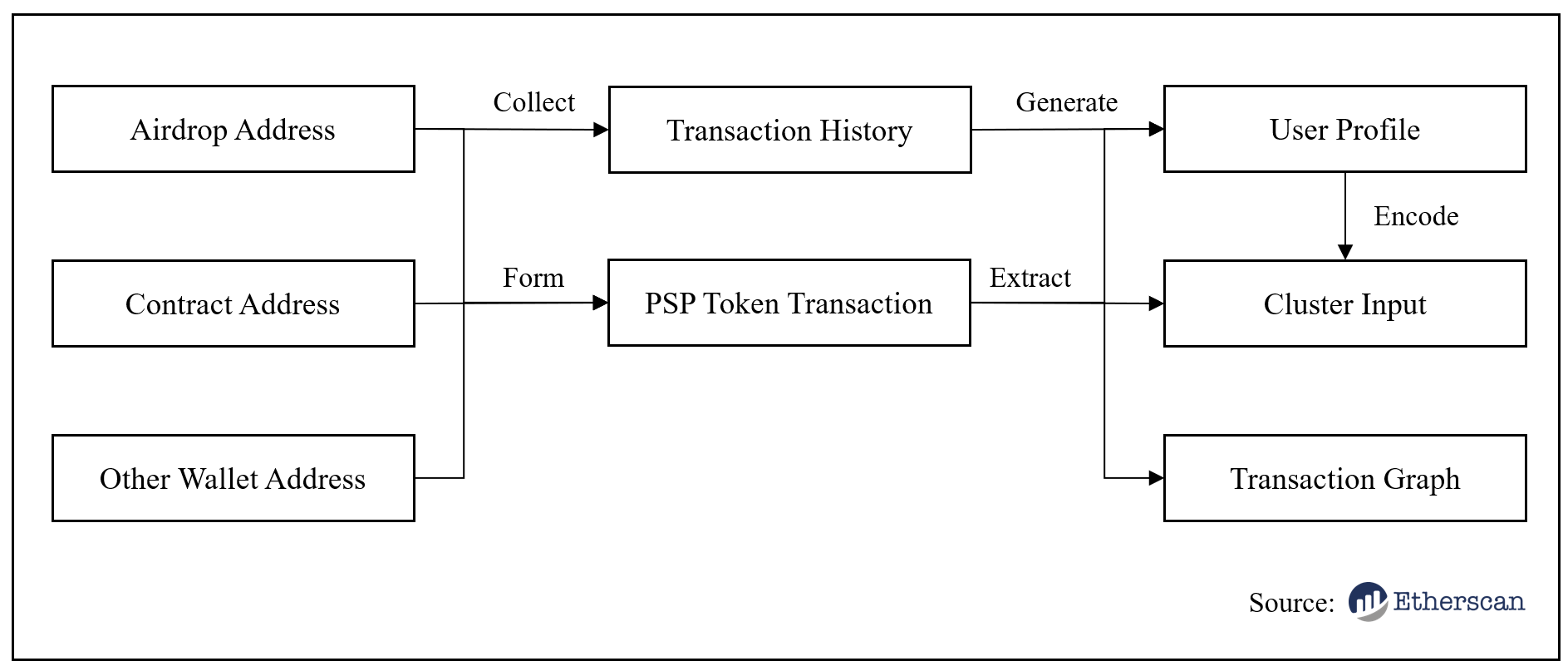}
    \caption{Procedure of Data Collection and Dataset Generate}
    \label{fig:data_collection}
    \Description{It is a flowchart of our data collection process, and the text near it provides equivalent information in terms of expressiveness.}
\end{figure}

In this paper, we collect 71,002 transaction records in the ParaSwap community formed by 20,148 members, among which there are 13,830 initial members. The general process of our data collection is shown in Figure \ref{fig:data_collection}. Since the raw on-chain data can be understood as bank transfer records, which are low in data dimension and contain less valuable information, we spend extra effort on data augmentation and pre-processing, such as compiling transaction's bytecode and interpreting the purpose of the function called. The dataset is formed by the behaviors of three kinds of roles: smart contract, initial and later members, among which the initial members are what interested us most. So, we collect the entire transaction history of these addresses, and together with the PSP token records, we generate a user profile and encode their behavioral patterns into a cluster-solvable sequence form in the following section. Moreover, we set up two transaction graphs, \emph{PSP Transaction Graph} and \emph{External Graph}, aiming to unravel the occurrence of cliques and the relationship between nodes.

\subsection{Behaviors after Receiving Rewards}
\label{sec:behavior_after_receive_reward}
\begin{table*}[ht]
\centering
\normalsize
\caption{Initial Member Behavior Statistics}
\label{tab:user_action_table}
    \begin{tabular}{|c|c|c|c|c|c|c|c|c|}
    \hline
    \textbf{Airdrop Type} &
      \textbf{Quota} &
      \textbf{\# Claimed} &
      \textbf{\% Selling} &
      \textbf{\% Buying} &
      \textbf{\% Staking} &
      \textbf{\% Sending} &
      \textbf{\% Receiving} &
      \textbf{\% LP}\\ \hline
    5,200 PSP   & 6,189 & 4,291 & 36.40\% & 0.02\% & 9.76\% & 31.76\% & 0.07\% & 0.05\%\\ \hline
    7,800 PSP   & 9,986   & 6,836 & 38.34\% & 0.09\% & 13.66\% & 19.18\% & 0.19\% & 0.10\%\\ \hline
    10,400 PSP  & 3,824    & 2,703    & 35.33\% & 0.26\% & 18.87\% & 11.95\% & 0.11\% & 0.11\%\\
     \hline
    \end{tabular}
\end{table*}

In order to explore the ParaSwap community member's behavior after the airdrop event, we have to categorize the transactions to simplify the data. In Web3, it is granted that operations on tokens can be traded on certain exchanges or direct transactions to an address. However, there are two operations with more financial properties that deserve more attention, namely \emph{Staking} and \emph{Liquidity Providing (LP)}. Staking is a kind of financial service held by DApps, which offers token holders a way of locking their assets to earn passive income without selling them. As staking tokens increases the number of locked positions, it reduces the liquidity in the market, which in turn can contribute to the stability and even appreciation of the coin price, which is positive for the members holding them. LP is also a kind of passive income and increases the value locked to stabilize the price. The tokens fill the DEXs' liquidity pools to make it easier for other traders to exchange tokens. In summary, from the community's perception, staking and LP are more likely to be perceived as actions that benefit their interests and, to a certain extent, can be considered as a form of altruism since the members performing these actions will sacrifice the liquidity of their assets and opportunity for quick cash.


Table \ref{tab:user_action_table} illustrates the distribution of transactions made by the initial members after the airdrop, based on three reward levels. It is noteworthy that the initial member who received 5,200 PSP tokens made a substantial proportion of the \emph{Sending} actions, however, the corresponding receiving transactions are comparatively low, regardless of the reward category. This suggests that a significant amount of PSP tokens were transferred to subsequent new members through direct transfers. A similar trend can be observed in the \emph{Selling} and \emph{Buying} behavior pair of trading.

Through simple calculation, we find that 103,745,200 PSP were claimed in total, from which subtract the remaining amount of PSP in initial members' addresses, we can get there are rounding 93,202,386 PSP flowed to the market, accounting for 89.34\% of all. Consequently, up to the moment of data collection on April 14$^{th}$, 2022, there are 11,948 initial members, accounting for 86. 39\% of all who gave away all their tokens and thus left the community. Among these people who withdrew, 3,282, 4,885, and 1,843 belong to the corresponding category of 5,200, 7,800, and 10,400. The attrition rates for the three groups were 76.49\%, 71.46\%, and 48.20\%.

\subsection{Most Popular Services}
\label{sec:data_extraction}

\begin{table}[ht]
\centering
\normalsize
\caption{Top 10 Contracts}
\label{tab:Top_10_contracts}
    \begin{tabular}{ccc}
    \hline
      \textbf{Name} &
      \textbf{Type} &
      \textbf{\# Times}\\ \hline
    Airdrop Contract & Airdrop & 13830\\
    ParaSwap   & Trading: swap & 3943\\
    Staking Pool 3    & Staking & 3400 \\
    Uniswap V3: PSP & Trading or LP & 2640 \\
    SushiSwap: PSP & Trading: swap & 1868 \\
    Staking Pool 4 & Staking & 1474 \\
    Uniswap V3: USDC-PSP & Trading or LP & 1452 \\
    Staking Pool 1 & Staking & 697 \\
    Airswap & Trading & 631\\
    Staking Pool 7 & Staking & 512 \\
     \hline
    \end{tabular}
\end{table}


We collect and organize a dictionary mapping 4,330 smart contract addresses with their names and functions to obtain a human-readable interpretation for each transaction. To explore what kinds of smart contract services have gained the most interaction in the wake of the airdrop event, we list the top ten smart contracts that received the highest number of transactions shown in Table \ref{tab:Top_10_contracts}. Not surprisingly, the vast majority of the smart contracts on the list belong to the type of financial services, as tokens from airdrop have not gained universal purchasing capacity in a short period.







\subsection{Network Construction}

\begin{figure}[htbp]
	\centering
    \includegraphics[width=0.8\columnwidth]{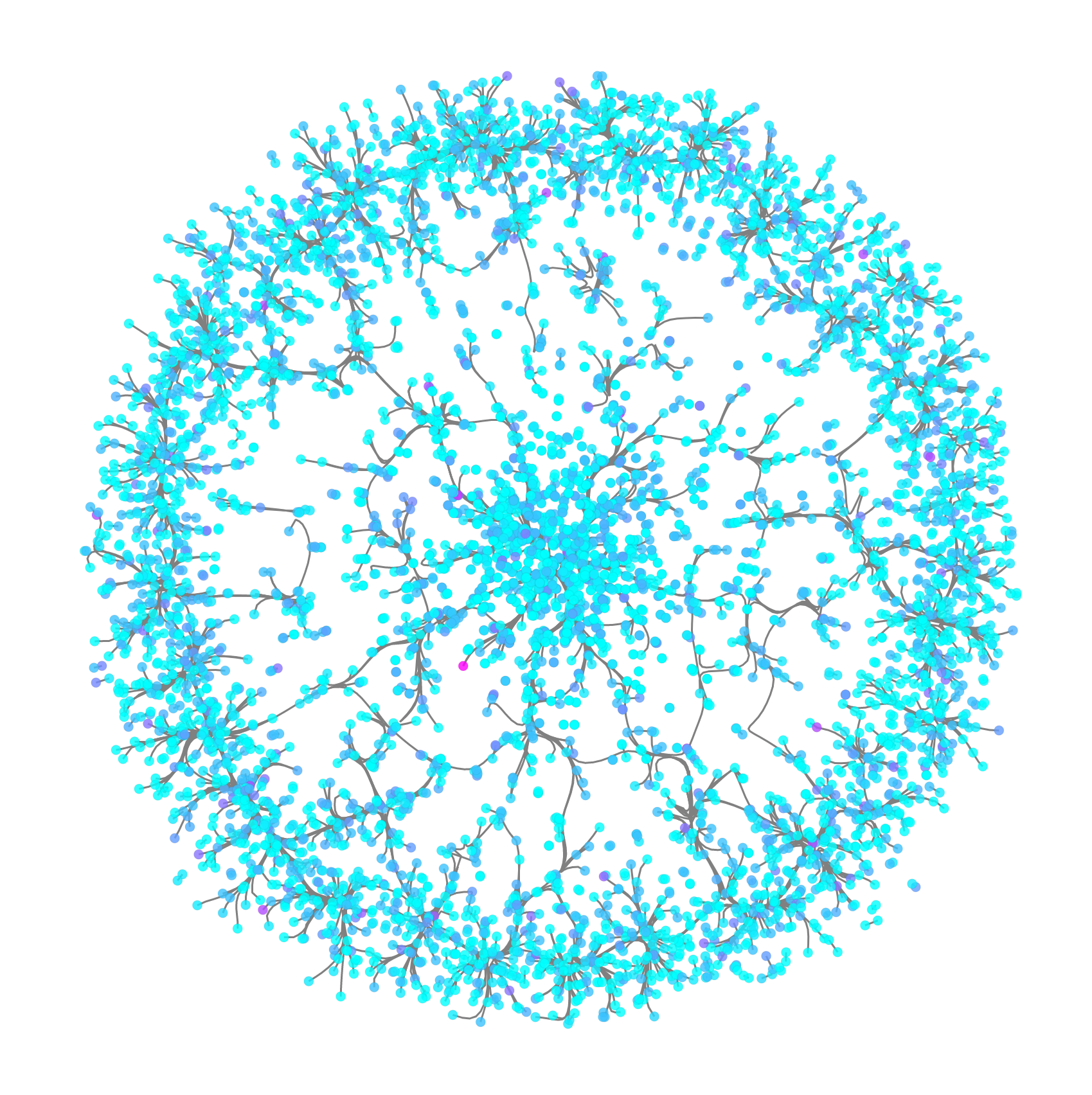}
    \caption{PSP Token Transaction Network}
    \label{fig:network_overview}
    \Description{It is a complex overview diagram showing the entire PSP Token network. It has a dense core and is surrounded by a Saturn-like halo. We use the blue to purple colormap to indicate how many of a particular node.}
\end{figure}

Based on the data obtained above, we collect the information about ParaSwap members' transaction records on Ethereum and create a token network, \emph{PSP Transaction Graph}, to perform temporal analysis, as shown in Figure \ref{fig:network_overview}. Specifically, we construct a directed graph for PSP tokens, where nodes stand for accounts and edges with weights representing transaction values. The network consists of 20,148 nodes and 46,410 edges for the PSP token transaction after the airdrop event.

\label{sec:network_construction}

In addition, we are also interested in what happened before the airdrop event. Based on the full external transaction records we collected, we generate an \emph{External Graph} containing 221,864 nodes and 1,111,209 edges for 7,003,317 external transaction history. Since an external transaction usually means a simple p2p transaction, it is not as informative as the PSP transaction graph, but it has a much larger and broader set of messages that can play an important role in exploring the connection between addresses before the airdrop.

\subsection{Temporal Evolution of Network}
\label{temporal_evolution_of_network}



\begin{figure}[hbp]
	\centering
    \includegraphics[width=\columnwidth]{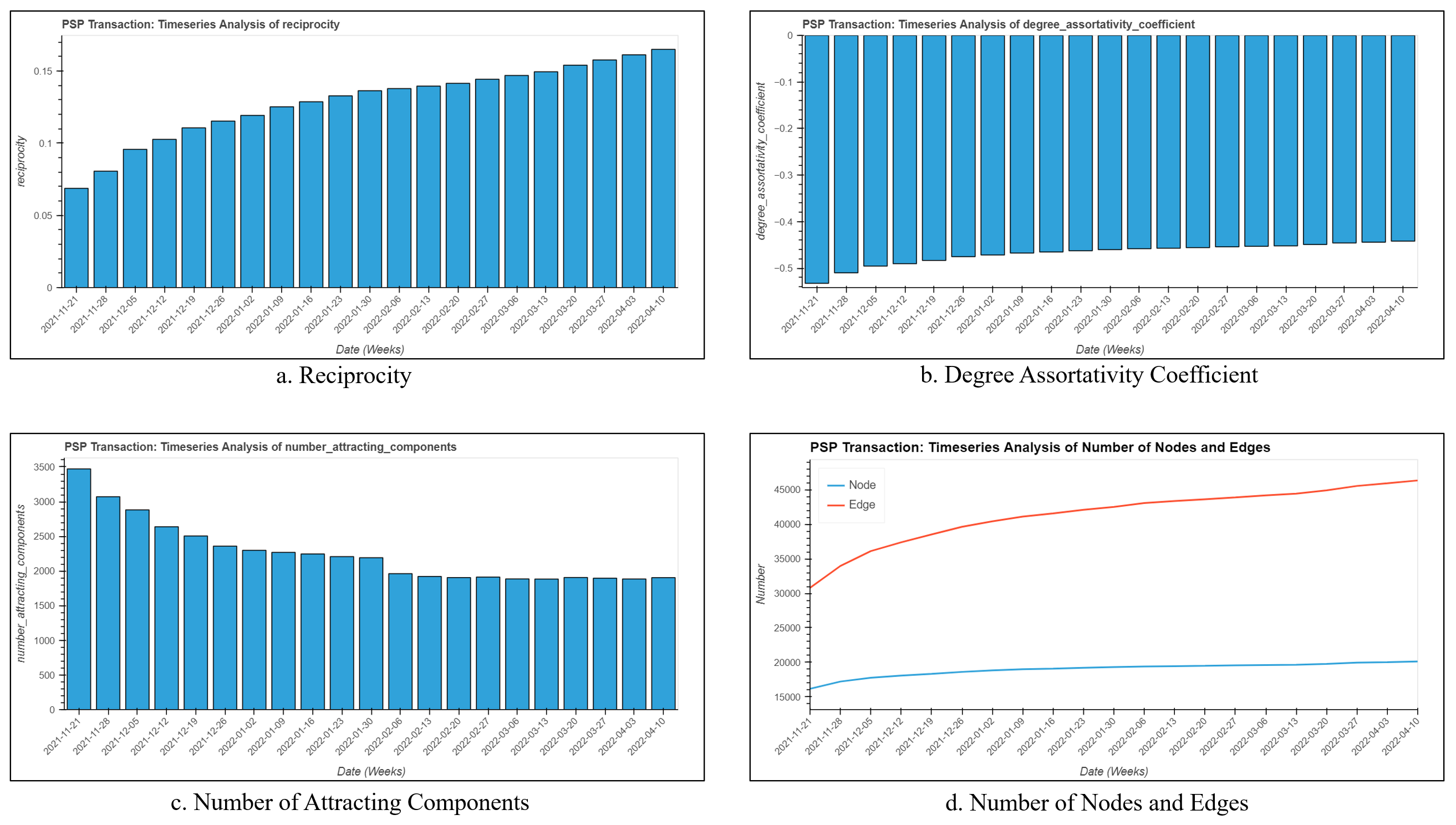}
    \caption{Selected Parameters of PSP Token Transaction Network}
    \label{fig:graph_overall_params}
    \Description{It is composed of four subgraphs, each of which has an x-axis with coordinates starting from November 21, 2021 to April 10, 2022. The subplots are labeled with alphabet. a is the reciprocity, which is a series of bars rising slowly from approximately 0.07 to 0.17. Similarly, b is the bar representing the degree assortativity coefficient, which rises from -0.5 to -0.44. c is the bar representing the number of the number of Edges is much higher than the number of Nodes and has a more pronounced upward trend. d represents that the number of the node and edge gradually increase at the beginning and stabilise over time.}
\end{figure}

As shown in Figure \ref{fig:graph_overall_params}, we compute selected properties by taking the slice by the week, which means that we intercept the graph from the origin to the moment of data collection every other week, including:
\begin{itemize}
    \item \textbf{a. Reciprocity:} Reciprocity originates from a social psychology concept that one positive behavior is rewarded by the performance of another positive behavior, which is defined as one of the defining attributes of any community \cite{wellman2018net}. The friendliness of others leads to friendlier responses and more willing to cooperate. Graph theory, refers to the likelihood of nodes in a directed network to be mutually linked.
    \item \textbf{b. Degree Assortativity Coefficient:} Assortativity is employed to assess whether nodes that share similar degree values tend to interconnect \cite{vasques2020transitivity}. A positive correlation in network assortativity is observed when nodes with higher degrees tend to associate with nodes that have a comparably large degree. If nodes with greater degrees tend to link to nodes with a lower degree, the network is said to have a negative correlation.
    \item \textbf{c. Number of Attracting Components:} An attracting component in a directed graph G is a strongly connected component with the property that a random walker on the graph will never leave the component once it enters the component, while the term strongly connected refers to if every vertex is reachable from every other vertex.
\end{itemize}

From the reciprocity (Figure \ref{fig:graph_overall_params}-\textit{a}), it can be seen that at the outset, there was only a one-way distribution of rewards to participants from the airdrop address, but over time the reciprocity gradually rises. As mentioned in the sociological explanation above, the whole network of transactions and the participants' connections in it are becoming more stable and robust. Due to the presence of exchanges, the degree assortativity coefficient (Figure \ref{fig:graph_overall_params}-\textit{b}) of this network seems to be unsurprisingly negative, implying a high degree of centralization. However, this negative value tends to decrease slightly over time, which can be explained by the fact that p2p transfers are increasing. In other words, there is an increased tendency for a node to link to another node with the same level of degree as itself. Again as the entire network began with airdrop contracts and moved from being highly centralized to relatively decentralized, the number of attracting components (Figure \ref{fig:graph_overall_params}-\textit{c}) also decreased over time because as PSP token circulates, isolates and non-airdrop addresses continue to join the market. The growth of the node and edge in the graph shows that they both gradually increase at the beginning and stabilise over time (Figure \ref{fig:graph_overall_params}-\textit{d}).



\section{Method for Role Identification}\label{sec:method_for_role_identification}

With the advancements in artificial intelligence, deep learning has become a commonly used method to understand user behavior by analyzing their interaction data \cite{10.1145/3097983.3098061}. However, this approach, despite being widely applied and requiring minimal domain knowledge, is not very interpretable. In particular, there is a lack of clarity regarding the discovery of roles and the understanding of user behavior within the "black box" of deep learning. For blockchain data, predictions can be made to identify attacks or money laundering \cite{MoneyLaundering}, but there is currently no widely recognized training set to guide the behavioral patterns that emerge from the data-level features. As an alternative to deep learning, some studies have explored the use of unsupervised learning or principal component analysis to obtain interpretable clusters, which can be used to more effectively capture and analyze the behavior of community members \cite{abdi2010principal}.

\subsection{Labeling Member's Behavior}
\label{sec:determining_the_user_activity}
\begin{figure}[htbp]
	\centering
    \includegraphics[width=\columnwidth]{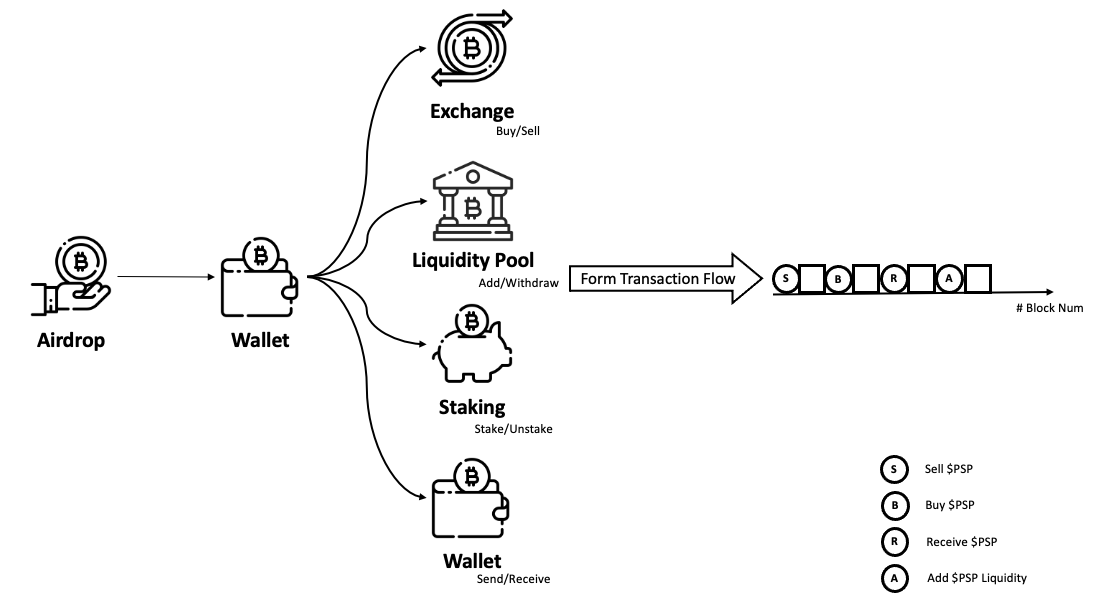}
    \caption{The Process of Formatting Transaction Flow}
    \label{fig:Transaction_Flow}
    \Description{Figure 5 is what can be considered as a visualization of our text about the feature extraction section.}
\end{figure}

For each address, we rearrange its activities into a transaction flow: a discrete sequence of events, which describes the smart contracts called by it, the number of PSP involved in the operations, and the remaining PSP in its balance after the operation. We divide the interactions between nodes into four categories: \emph{Trading}, \emph{LP}, \emph{Staking} and \emph{Transferring}. Figure \ref{fig:Transaction_Flow} illustrates the formatting of a transaction flow.


\subsection{Feature Extraction}
\label{sec:feature_extraction}
Our clustering algorithm is based on a similarity graph, where each node represents an address, and each edge represents a similarity weight between two addresses' transaction flows. We identify the address behavioral clusters by partitioning the similarity graph. 

We extract transaction patterns from the flow as features to compare similarities. Specifically, we extract operation types of the $m^{th}$ address and reformulate an operation flow as a sequence $S_{m} = \{s_1, s_2, \dots, s_q\}$, where $s_i$ is the $i^{th}$ operation in this sequence, and $q$ is the total number of items in the sequence. We use $\mathcal S$ to denote the set of all sequences and $M$ to represent the total number of sequences in $\mathcal S$. However, the difference in the length of operation flows make it difficult to measure the similarity. To solve the above problem, we denote eight kinds of operations with $j$ as the index. After calculating and organizing the number of operation types in every address, we can obtain the features of the $m^{th}$ address as $O_{m} = \left\{c_1, c_2, \dots, c_8\right\}$ , where $c_j$ equals to 1 or 0. When it is equal to 1, the address has done this operation. When it equals 0, this operation has not been performed by the address. After extracting the features of addresses' behaviors, we choose weighted cosine similarity over other alternatives (e.g., Euclidean distance) because it is more suitable to handle the highly sparse matrix \cite{houle2010can}.

\begin{equation}
    d(S_1, S_2) = \frac{\sum_{j = 1}^{n} (\omega_j c_{1, j}) \times (\omega_j c_{2, j})}{\sqrt{\sum_{j = 1}^{n} (\omega_j c_{1, j}^{2})} \times \sqrt{\sum_{j = 1}^{n} (\omega_j c_{2, j}^{2})}}
\end{equation}

$d(S_1, S_2)$ ranges from 0 to 1, and a small distance value indicates a high similarity between two transaction flows. $\omega_j$ refers to the weight of the $j^{th}$ feature, whose value chosen in our method is fundamental to the cluster results. $n$ is the number of elements in $O_m$, which is $8$ in this case. 

\subsection{Clustering Algorithm}
\label{sec:cluster_algorithm}
We employed an agglomerative hierarchical clustering (AHC), which has less time complexity and better computational stability \cite{zhou2016method}. Specifically, AHC procedures start with $m$ singleton clusters and repeatedly merge clusters until a specific level of granularity has been achieved, i.e., the number of clusters reaches the target. The AHC has three main distance measures: single linkage, complete linkage, and average linkage \cite{zhou2016method}. In most cases, the AHC with single link is relatively stable and effective for linear, manifold, and convex data. Another important issue is how to obtain the suitable partition from several clustering results, which can be characterized as the problem of determining the optimal number of clusters $K$.



\subsection{Determining the Number of Roles}
\label{sec:determining_the_number_of_roles}

In our approach, the number of clusters, $K$, is a free variable, and how to decide its value will be the key to the whole process. Too high or too low a value of $K$ will have a significant impact on the results, especially in terms of interpretability. To overcome the problem, we used the Silhouette Coefficient \cite{Silhouette_Coefficient} to select the optimal $K$ in AHC . Specifically, we trained the data on AHC with $K$ ranging from 2 to 20 to determine the optimal number of clusters. We find that the AHC with $K \in \left[12, 20\right]$ seems to be a good option in this circumstance. 



After calculating the transaction distribution for each cluster and comparing its differentiation from other $K$ value results, we find that the most suitable solution is when $K$ = 14, which results in a relatively balanced number of addresses among the clusters, but at the same time maintains independence between, without fusing important and distinct characteristics.

\subsection{Cluster Results}

\begin{figure}[htbp]
	\centering
    \includegraphics[width=\columnwidth]{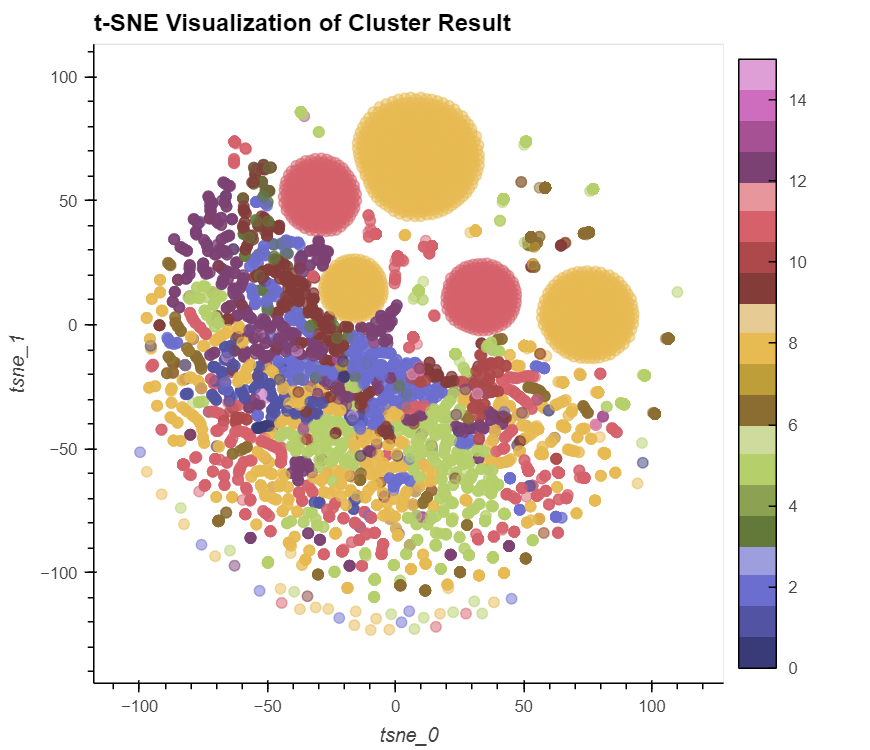}
    \caption{t-SNE Processed Visualization of Cluster Results}
    \label{fig:t-SNE}
    \Description{Figure 6 shows the t-SNE down-scaled visualization of the clustering results, which has 14 colors to represent the different clusters, where the large red and yellow clusters can be clearly seen. There are also other small groups in the periphery composed of various colors.}
\end{figure}

After determining the parameter, we can proceed with the model to get the initial members' roles in the airdrop community. The result generated 14 clusters, and the t-Distributed Stochastic Neighbor Embedding (t-SNE) visualization is shown in Figure \ref{fig:t-SNE}.

In Table \ref{tab:cluster_results}, we label the clusters with an index and description of their proportion, source of income, and ways of consumption. It is worth noting that the algorithm captures the type of operation of the members, and the ways of consumption describe precisely their behaviors belonging to the corresponding cluster with very clear boundaries. This means that 75.97\% of the members performed a single type of operation after the airdrop event. Data from transaction flow shows that most of these operations were done once, which is consistent with the context of high Ethereum commission fees. As part of the single-operation clusters, 61.03\%, or clusters 1 and 2, exited the community in a short period after obtaining an airdrop, which is also validated and in accordance with the member attrition scenario we discussed in Section \ref{sec:behavior_after_receive_reward}.

\begin{table}[htbp]
\centering
\caption{14 Clusters of Members in ParaSwap Community}
\label{tab:cluster_results}
\begin{tabular}{|c|c|c|c|}
\hline
\textbf{Index} & \textbf{Source}                                                                    & \textbf{Consumption}    & \textbf{Percentage} \\ \hline
1              & \multirow{11}{*}{\begin{tabular}[c]{@{}c@{}}Airdrop\\ (Group 1)\end{tabular}}      & Selling                 & 38.79\%             \\ \cline{1-1} \cline{3-4} 
2              &                                                                                    & Sending                 & 22.24\%             \\ \cline{1-1} \cline{3-4} 
3              &                                                                                    & Staking                 & 14.94\%             \\ \cline{1-1} \cline{3-4} 
4              &                                                                                    & Staking+Selling         & 8.50\%              \\ \cline{1-1} \cline{3-4} 
5              &                                                                                    & Holding                 & 4.87\%              \\ \cline{1-1} \cline{3-4} 
6              &                                                                                    & Selling+Sending         & 2.39\%              \\ \cline{1-1} \cline{3-4} 
7              &                                                                                    & Staking+Sending         & 1.71\%              \\ \cline{1-1} \cline{3-4} 
8              &                                                                                    & LP+Staking              & 0.39\%              \\ \cline{1-1} \cline{3-4} 
9              &                                                                                    & LP+Selling              & 0.30\%              \\ \cline{1-1} \cline{3-4} 
10             &                                                                                    & LP                      & 0.11\%              \\ \cline{1-1} \cline{3-4} 
11             &                                                                                    & LP+Sending              & 0.09\%              \\ \hline
12             & \multirow{3}{*}{\begin{tabular}[c]{@{}c@{}}Airdrop + Buy\\ (Group 2)\end{tabular}} & Staking+Sending/Selling & 3.58\%              \\ \cline{1-1} \cline{3-4} 
13             &                                                                                    & Sending/Selling         & 1.91\%              \\ \cline{1-1} \cline{3-4} 
14             &                                                                                    & LP+Sending/Selling      & 0.18\%              \\ \hline
\end{tabular}
\end{table}

\subsection{Composition of Airdrop Tiers in Clusters}

\begin{figure}[htbp]
    \centering
    \includegraphics[width=\columnwidth]{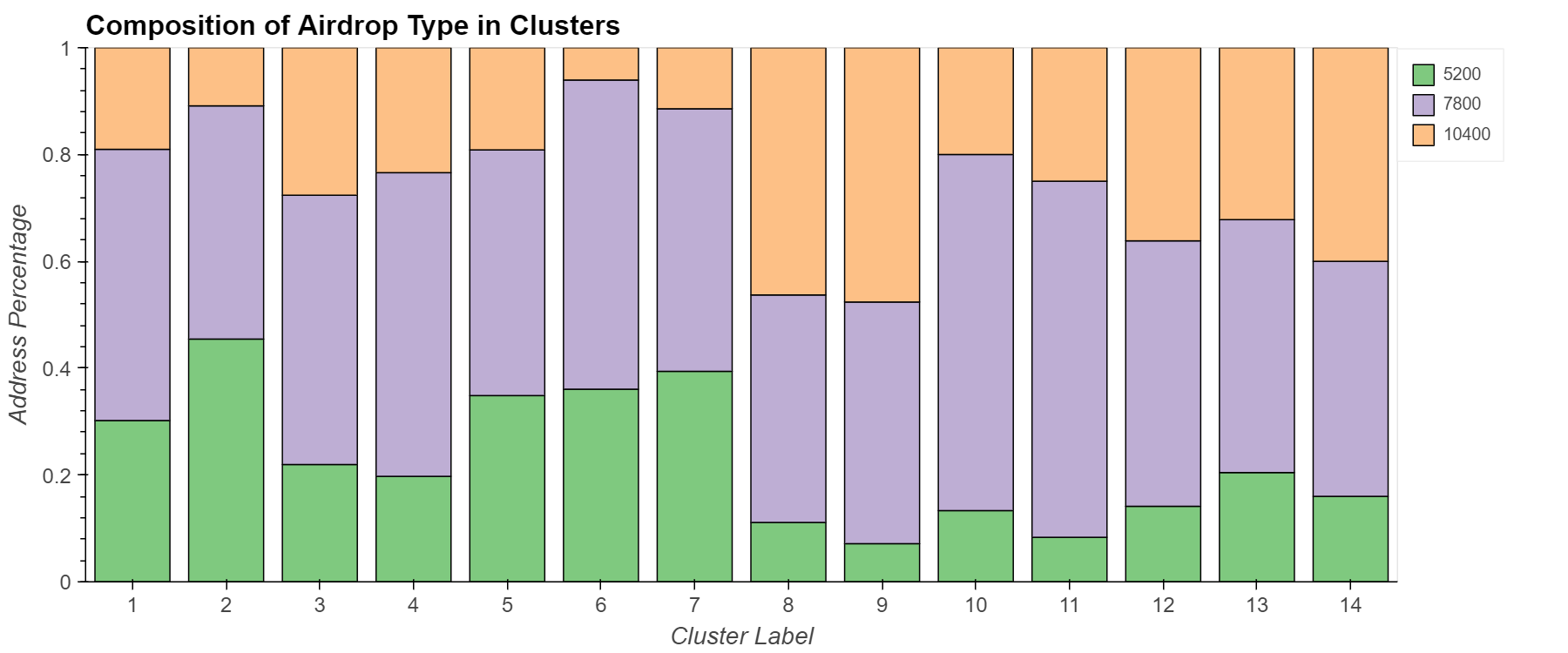}
    \caption{Proportion of Airdrop Tiers in Clusters}
    \label{fig:proportion_of_airdrop_tiers_in_clusters}
    \Description{Figure 7 is a stacked histogram. x-coordinates are the 14 clusters. all x's sum to 1 in the y-direction to reflect the constituent components. We use green, purple and orange to represent the airdrop reward levels of 5200, 7800 and 10400, respectively. In all clusters, purple occupies the majority. However, there are some special clusters such as 7 or 8 where orange dominates.}
\end{figure}

By calculating the proportion of reward recipients at different tier levels in each cluster, we obtain Figure \ref{fig:proportion_of_airdrop_tiers_in_clusters}. On the whole, addresses belonging to the 7,800 tier occupy a significant majority in each cluster, which is inevitable, as the number of claimants for the three-tier rewards of 5,200, 7,800, and 10,400 are, respectively, a ratio close to \emph{3 : 5 : 2}. 

Taking the above factors into consideration, we can say that members with proactive token purchasing behavior are statistically more likely to be involved in more permanent and substantial financial activity. Initial members with these characteristics are what DApps want to see, as their active participation in staking activity enhances the stability of the issued tokens. However, this segment accounts for less than 6\% of the total addresses eligible for the airdrop. How to increase the percentage of these addresses should be of interest and research to project owners.

\subsection{Implications for Buying Behavior}
\label{sec:implications_for_buying_behavior}
We define the operation of exchanging PSP from a DEX as a buying behavior, which means that these members in addition to airdrop rewards, in making the act of collecting tokens with subjective motivation. So, naturally, the research questions such as "What are the qualities of an address that has a buying behavior?", "How does this group of buyers differ from other groups?" arise.

Shown in Table \ref{tab:cluster_results}, we mark the cluster index from 1 to 11 as \emph{Group 1}, whose only source of PSP token is airdrop, and the index from 12 to 14 as \emph{Group 2}, whose performed buying behaviors. We selected three indicators: \textit{Balance}, \textit{Staking}, and \textit{Liquidity} to discuss the distinction between the two groups in terms of duration period, and quantity held. We define the sum of the number of days that a financial activity is carried out by an address until the complete withdrawal of capital as \textit{Period}, while the weighted average of the number of tokens corresponding to the duration is defined as \textit{Quantity}, which in this case is of course number of PSP holding.

\begin{figure*}[htbp]
	\centering
    \includegraphics[width=1\textwidth]{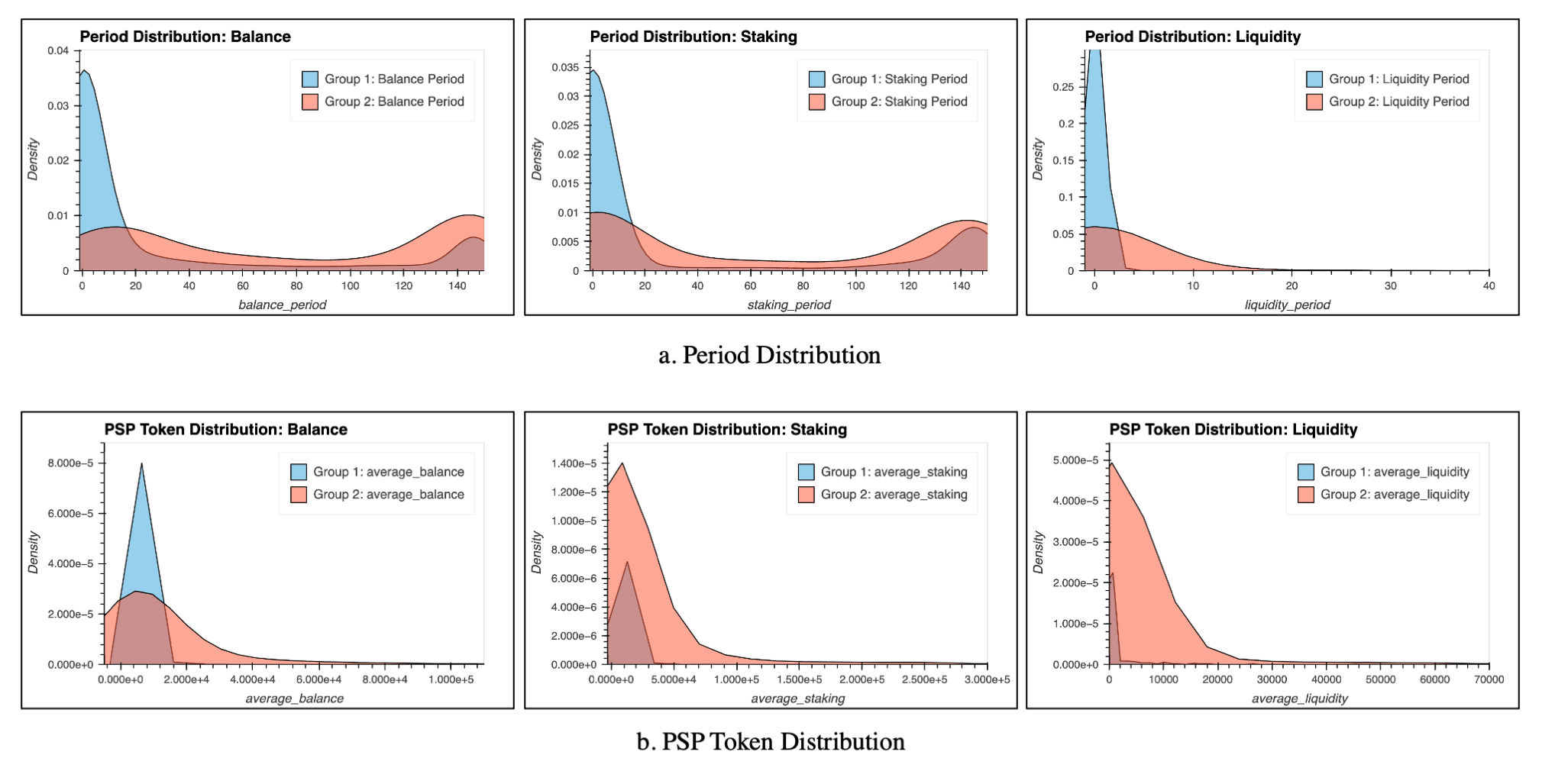}
    \caption{Fitted Period Distribution and PSP Token Distribution with PDF}
    \label{fig:distribution_groups}
    \Description{Figure 8 is composed of two sets of probability density function plots, each with three plots representing balance, staking and LP. a group is the PERIOD distribution, depicting the duration of investment of the assets being used for the three purposes. x axis is time, in days. For all uses, the PDF for Group1 is clustered around 0 and the curve is smoother. b group is the PDF on the amount invested. the smaller sample size due to the fact that part of the Group1 sample is not involved in staking and liquidity leads to a more jagged curve. Group2 has a broader distribution on the x-axis. b group x-axis upper limits are 10000, 30000, 70000.}
\end{figure*}

In Figure \ref{fig:distribution_groups} \textit{a}, we fit their period distributions with probability density function (PDF) in order to reduce the gap between \textit{Group 1} and \textit{Group 2} in terms of the amount of data. The overall perspective of the total density distribution shows that \textit{Group 2} has a wider distribution than \textit{Group 1}. In the leftmost Balance Period, there is a relatively clear concentration at 0 and 147 on the x-axis, meaning that initial members are more likely to remove their PSP tokens on the day they receive a short or to continuously hold it until the day we made a slice to collect the data. It is worth noting that in the rightmost liquidity period distribution, the maximum duration value is just under 40, which means that for both groups, overall, all members are not very willing to provide liquidity.

In Figure \ref{fig:distribution_groups} \textit{b}, we plot the density functions for the quantity of tokens of \textit{Group 1} and \textit{Group 2}. There is a clear difference between the three subgraphs here and those above: the top graph PDF curve is smoother, while the bottom one is more serrated, which indicates that the Figure \ref{fig:distribution_groups} \textit{b} has a more discrete distribution of values, caused precisely by the fixed amount of three airdrop slots. Starting with the distribution of balances on the left, there is no doubt that \textit{Group 2}, which has "purchases" as an additional source of PSP tokens, has more members distributed over a higher quantity of tokens. As for the middle graph of staking and right graph of liquidity, \textit{Group 2} also outperformed \textit{Group 1} in both respects, with the majority of members in \textit{Group} 1 having no staking or LP action at all, which results in missing data. The only samples stay low and concentrated around the origin.

\subsection{Discussion on Role Identification}
\label{sec:discussion_on_role_identification}
Fourteen clusters are, as yet, way too many for conducting higher-level analysis and discussion. Coupled with the fact that we noticed that clusters are logically connected with each other in actions beyond the data level, we re-integrated the clusters into five roles. We next elaborate on the rationale for the reintegration and will discuss the instructive significance of the roles and classification methods in the next section.

\subsubsection{Speculators (41.18\%): Cluster 1, 6}
Speculators will choose to sell the airdrops through CEXs or DEXs without any other operations after obtaining the airdrops rewards. The motivation behind their almost immediate withdrawal from the community was most likely purely profit-driven. Speculators may have given a return that was just able to cover their costs when they participated in the preliminary activities. Based on their judgment, selling PSP in the short term would yield the greatest return, which also implies that they hold a pessimistic attitude about the future development of the agreement. Otherwise, promising development would bring them higher expected returns.

So, can we consider the presence of speculators, especially as initial members, to impede the development of the Web3 community? We suggest that the answer is no. There are plenty of speculators in whatever market, and even more so in the context of blockchain, known for its free trade. This should be treated as an immutable environment precondition. Speculators are bringing in new members as they exit the community. But DApps need to be wary that once the number of speculators becomes too large and the second and third-generation members all hold this pattern of behavior, it will lead to the ponziization of the protocol and bring about a difficult reversal of the predicament \cite{bartoletti2020dissecting}.

\subsubsection{Diamond Holders (20.31\%): Cluster 3, 5, 8, 10}
Diamond Holders are those initial members who hold PSP until April 13$^{th}$ 2022 without selling or sending PSP. They are considered optimistic about the long-term development of the community in terms of the appreciation of token price, hoping to achieve greater income in the later stage. We further divide two sub-categories in Diamond Holders:
\begin{itemize}
    \item \emph{Risk-averse Holders} (4.87\%): They are risk averse, refusing to participate in the risky activities associated with tokens, such as staking and LP, despite the potential income.
    \item \emph{Risk-seeking Holders} (15.44\%): Different from risk-averse holders, risk-seeking holders are keen to participate in risky activates.
\end{itemize}

Diamond Holders are undoubtedly one of the most welcoming categories of community members. They are willing to give up quick cash and bear the opportunity cost\footnote{The concept of opportunity cost is fundamental to the economist's view of costs. Since resources are scarce relative to needs, the use of resources in one way prevents their use in other ways.} of holding positions and contributing their assets to the PSP price. In contrast to speculators, they have confidence and are qualified to participate in the future voting of the community for holding governance tokens.

\subsubsection{Airdrop Hunters (22.24\%): Cluster 2} We suggest that the vast majority of the initial members belonging to cluster 2 should be affiliated accounts controlled by a single individual, since the risk of p2p trading with strangers is often unacceptable without smart contracts as intermediaries rather than direct transaction. We will prove this hypothesis later in Section \ref{sec:arbitrager_cliques}.

\subsubsection{Diversified Members (10.62\%): Cluster 4, 7, 9, 11}
Different from speculators and airdrop hunters, Diversified Members choose to take part in multiple activities. They are involved in staking and LP, but at the same time, they do send and sell. Members of this composite behavior pattern are more likely to be independent individuals, as savvy Airdrop Hunters tend to aggregate PSP in one place before staking or LP to reduce multiple commission fees.

On the one hand, they pursue more returns by participating in staking and LP. On the other hand, they also sell PSP for a profit when PSP reaches a proper price. Their decision patterns are relatively more dynamic and flexible. In a sophisticated decentralized community, this group of members should make up the majority, but we only find around ten percent of them in the ParaSwap community.

\subsubsection{Buyers (5.67\%): Cluster 12, 13, 14} Also welcomed by the community, but unlike Diamond Holders, Buyers are more proactive in purchasing PSP. As we mentioned in Section \ref{sec:implications_for_buying_behavior} for positive implications of purchasing behavior, we classify this group independently as a role.

\subsubsection{Summary}
To put a summary at the end of this section, we would like to discuss the potential contribution of role identification to decentralized community governance and airdrop distribution methods as a practical guide, which can also be used as a universal methodology to understand the human engagement within other blockchain based community.

We found that the differential airdrop can influence the behavior of initial members to a degree. Members who receive more rewards have more spare capacity to invest in positive actions for the community. We cannot conduct a profound analysis on the cost of acquiring airdrop per initial member in this paper because it involves extremely complex calculations and system design. If rewards can fill the cost of early-stage participation, members will have fewer concerns about staking and LP. A more accurate prediction when setting the differential regulations, designed to ensure that members do not lose money without affecting the stability of the token model, would provide a stronger incentive.

On a more assertive side, the community could split the planned airdrop into multiple installments or set additional rewards. Specifically, the protocol can perform role identification with respect to the behavior of the address after the first few airdrops, during which they investigate the role dynamics of members, including how their role changes over the cycles and what kind of factors can preserve long-term participation in the community. The project owner can decide on the airdrop list through the above process and result. While such a policy is somewhat paternalistic, and we would like to see more spontaneous enthusiasts, on a result level, the tempting additional rewards could lead to more positive behavior. Community rule makers also should not have the illusion of being overly idealistic; a profit-oriented mindset is more prevalent on anonymous and highly market-oriented blockchain platforms than in real communities \cite{article:anonymity_1, article:anonymity_2}.

\section{Network Analysis}
\label{sec:component_analysis}

We try to explore the behavior of the initial members through component analysis on \emph{PSP Transaction Graph}. The existence of CEXs will cause great confusion in the transaction analysis, making finding components extremely difficult. The exclusion of CEXs may cut down on the number of components, but could provide a more credible dataset. Hence, we focus on those components that consist entirely of direct transfers in order to obtain more accurate and credible results. Through the process of cutting off contract addresses and isolates, we obtain 2,493 components consisting purely of 7,982 wallet addresses and label them with corresponding IDs.

\begin{figure}[htbp]
	\centering
    \includegraphics[width=\columnwidth]{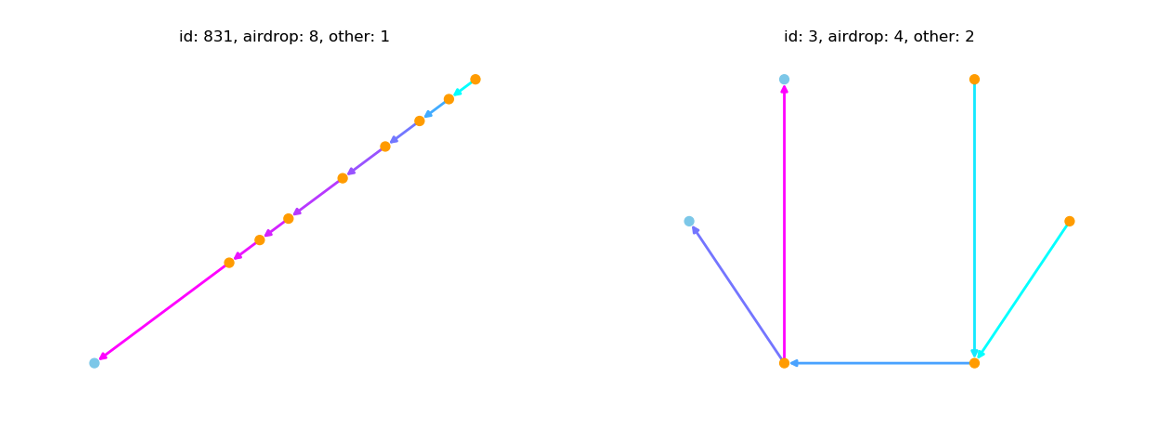}
    \caption{Typical Structure of Chain Transaction}
    \label{fig:transaction_flow_chain}
    \Description{The left side of Figure 9 is a series of points connected like a chain, and the edges connecting them are unidirectional and progressively deeper, which means that the funds are accumulating in the transfer. On the right side is a dumbbell-like shape of the dots, starting with two initial members passing funds to two second-generation members in a unidirectional manner.}
\end{figure} 

\subsection{Typical Forms of Component}
\label{sec:typical_forms_of_component}

We found a large number of records of p2p transactions in Section \ref{sec:determining_the_number_of_roles}, so how these PSP were transferred became a natural question. For 2,493 components with marked IDs, we recorded the number of nodes and edges as well as the edges' weight, the number of initial and newly-generated members, and reciprocity characteristics to produce a profiling dataset. On average, each component consists of 3.20 nodes, and there are 586 components containing more than three. 3,219 out of 7,982 second-generation members are involved in these p2p transaction components. On average, each component contains at least 1.38 second-generation members, and only 35 of them do not contain any. When we plotted all the components containing more than three nodes, we noticed some special structures that frequently appeared and immediately associated them with potential arbitrage behavior. We named the two typical structures as \emph{Chain} and \emph{Sunflower} according to their propagation method. We use different node colors of \emph{Orange} to represent initial members and \emph{Sky Blue} to represent later members. In order to have a better visual perception of a white background, we use the \emph{cool} colormap array\footnote{https://www.mathworks.com/help/matlab/ref/cool.html}, which changes its color from blue to purple as the edge weight increases.

As shown in Figure \ref{fig:transaction_flow_chain}, we find two special paths of one-way token transfer among components, which we call the "\emph{Chain}" transaction. The pattern starts from one or multiple initial members and continuously propagates the PSP it received to its successors. As shown in the subgraph on the left of Figure \ref{fig:transaction_flow_chain}, in the component ID-831, eight nodes of initial members pass their tokens to a second-generation member. The edge color between the nodes is from shallow to deep, representing the increasing number of PSP passed like a Fibonacci series. The other one-way pass structure of ID-3 has more of a merging and splitting pattern, as shown on the right.

\begin{figure}[htbp]
	\centering
    \includegraphics[width=\columnwidth]{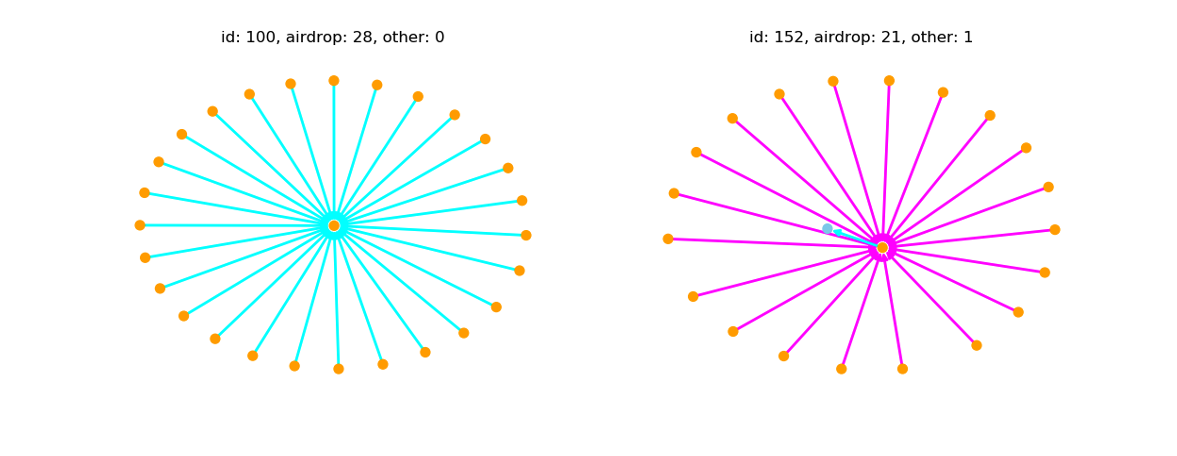}
    \caption{Typical Structure of Sunflower Transaction}
    \label{fig:transaction_flow_assemble}
    \Description{The two subgraphs in Figure 10 are similar. They are both radial aggregation structures like sunflowers or umbrellas. They both have an initial member at the center.}
\end{figure} 

Another transmitting pattern we called "\emph{Sunflower}" is shown in Figure \ref{fig:transaction_flow_assemble}. The PSP obtained from the initial members is transferred to an eligible address, which performs a single operation on the PSP. The difference between these two approaches is that the initial central member in the right-hand side of Figure \ref{fig:transaction_flow_assemble} acts as a transit vertex to transfer the collected PSP to another newly-generated member.

\begin{figure}[htbp]
	\centering
    \includegraphics[width=\columnwidth]{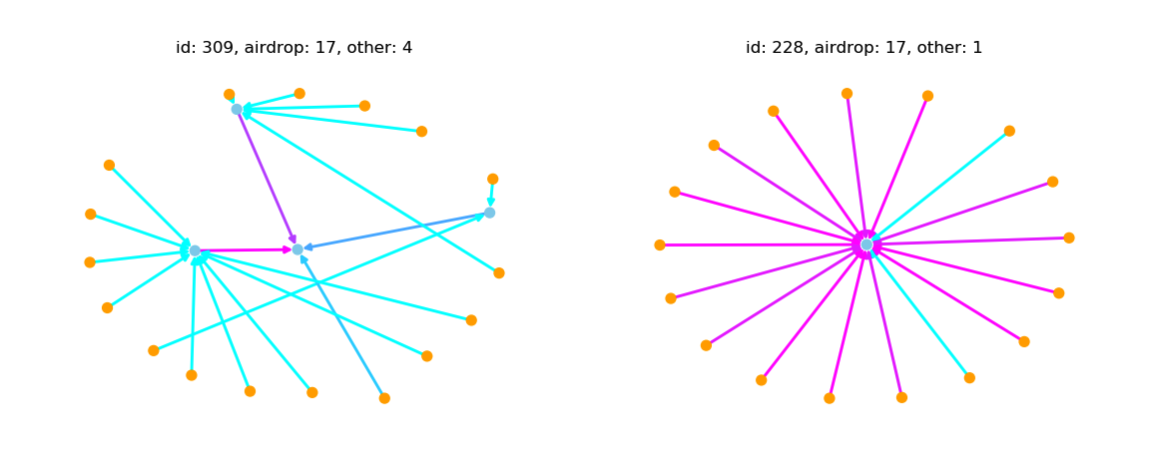}
    \caption{Alternative Structure of Sunflower Transaction}
    \label{fig:token_flow_assemble_2}
    \Description{The left side of Figure 11 also shows a radial convergence structure, but unlike Figure 9, it has some more nodes in the form of staging stations. The right side of Figure 10 is a standard sunflower structure, but unlike the previous one, it has a newly generated member in the center.}
\end{figure} 

The two patterns shown in Figure \ref{fig:token_flow_assemble_2} also aggregate PSP to a single node. But unlike the "\emph{Sunflower}" transaction, the central nodes are non-airdrop nodes. We dig into the transaction history of these central nodes and notice that they have no record of interaction with any smart contract but only serve as a staging area, which prevents the airdrop nodes from directly interacting with each other in order to ensure the independence between them. If without a network-wide screening of transactions, it would be difficult for the DApps to exclude such components.

Admittedly, discovering these components based on the PSP token network while proving the existence of potential arbitrage is a kind of hindsight. DApps would undoubtedly like to have discovered these possible arbitrage behaviors before the airdrop event. However, this does not diminish the contribution of this section. First of all, the emergence of this sort of aggregation patterns has demonstrated the vulnerability of the ParaSwap community's mechanism for filtering hunters. A component with a more significant number of members, such as ID-100, is capable of holding 130,000 PSP in a single individual and has more than 20 times the voting power of the average member. In addition, we demonstrate that component analysis remains essential even after the airdrop event, giving the community a way to track down potential hunters or Sybil attackers and eliminate the risk before they cause significant damage.

\begin{figure}[htbp]
	\centering
    \includegraphics[width=\columnwidth]{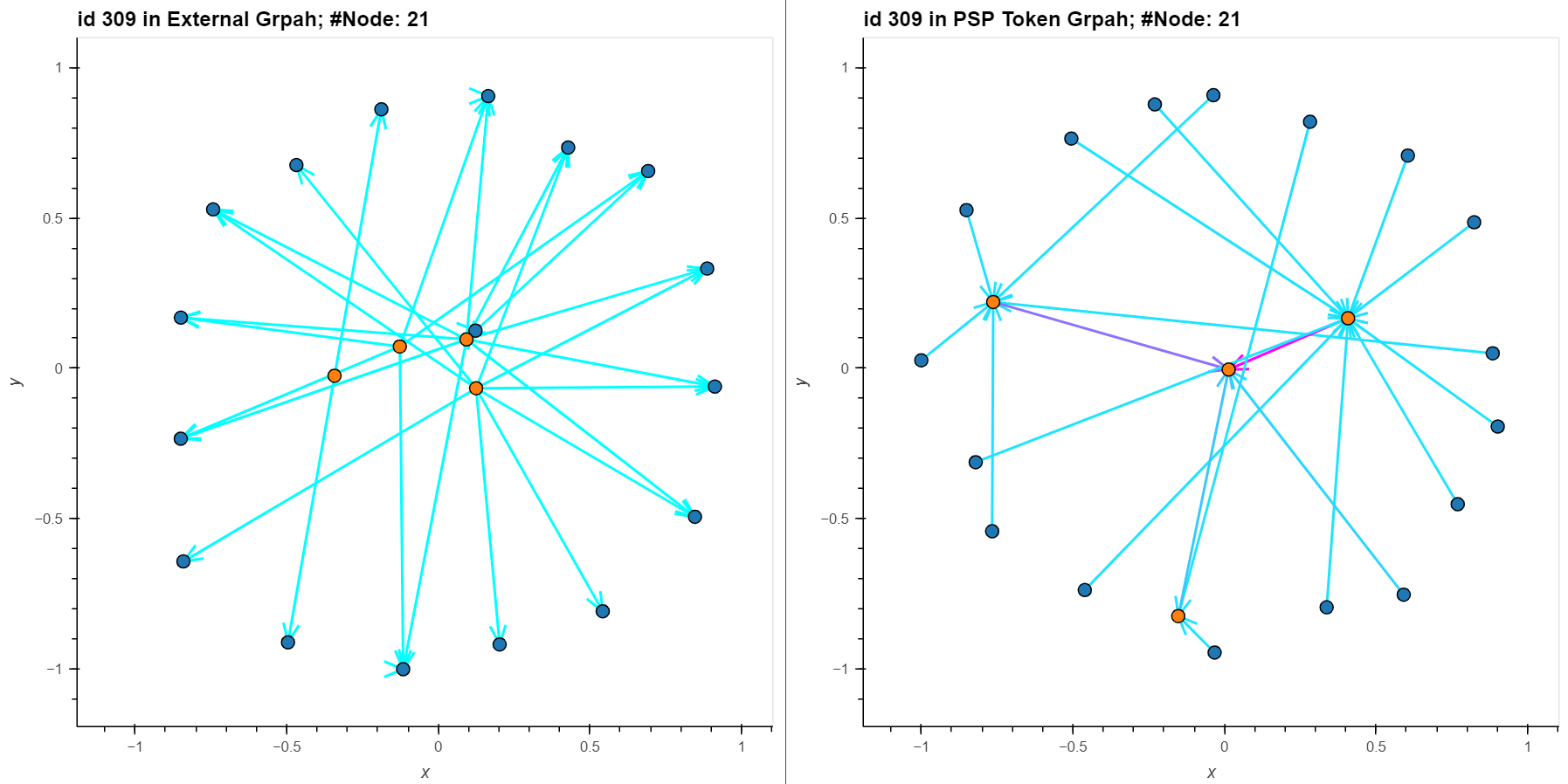}
    \caption{Clique ID-309}
    \label{fig:id-309}
    \Description{Figure 12 shows the network generated by ID-309 in the external graph and the psp transaction graph, respectively. the center of the structure in the external graph is the four funding points, and each node that goes to participate in the airdrop bid is supported by several funding nodes. In the psp transaction graph, the rewards received are returned to the funding nodes. More detailed description is in the text part of the paper.}
\end{figure}

\subsection{Airdrop Hunter Cliques}
\label{sec:arbitrager_cliques}
By comparing the external graph before the airdrop event with the PSP transaction graph after it, we attempt to find the hunters that are not allowed by the ParaSwap policy, where one individual or organization controls multiple addresses to obtain multiple rewards.

As mentioned in the above section, the presence of CEXs would substantially obfuscate the flow of tracking transactions, we choose to deal with the most intuitive and obvious p2p token transactions. We numerically assign each component with ID and exclude behaviors that have a low probability of being identified as arbitrage, such as those components consisting of only two or three members, and instead focus on subgraphs with more than five nodes, which is the threshold of getting ParaSwap's focus when examining eligibility\footnote{https://medium.com/paraswap/whats-an-active-user-clarifying-psp-token-distribution-filtering-logic-81df6096d410}. Our goal is to evaluate existing allocation methods to discuss their plausibility and discover what methods hunters use to escape scrutiny from DApps. To address this issue, we retrieve the members' information of components appearing in the PSP transaction graph and match them to the external graph to discover their interrelationship before detection.


\subsubsection{The Organized and Planned Hunter}
\label{sec:the_organized_and_planned_arbitrager}

This small clique's center consists of 4 non-airdrop addresses, as presented in Figure \ref{fig:id-309}. The other 17 initial members were not related in any way prior to the award, but they were all funded by these four "sponsors" through direct transfers to provide funds for airdrop eligibility. This sponsorship was sophisticated planning, with each initial member being backed up by multiple sponsors, and the PSP tokens were smoothly returned by the sponsorship route as designed after the rewards were distributed.

Given that registering addresses on the blockchain costs nothing, and since there is no direct connection between addresses participating in airdrop slots, community regulators can barely identify and decipher fake airdrop bidders' behavior generated by "Sponsorship Models" like this, unless they look further than their second degree of association. However, given social and public opinion considerations, a strict but imprecise screening policy could trigger more intense resistance from already disgruntled users, exposing the program to irreversible reputation damage and user loss.

\subsubsection{The Cautious Hunter}
\label{sec:the_cautious_arbitrager}
\begin{figure}[htbp]
	\centering
    \includegraphics[width=\columnwidth]{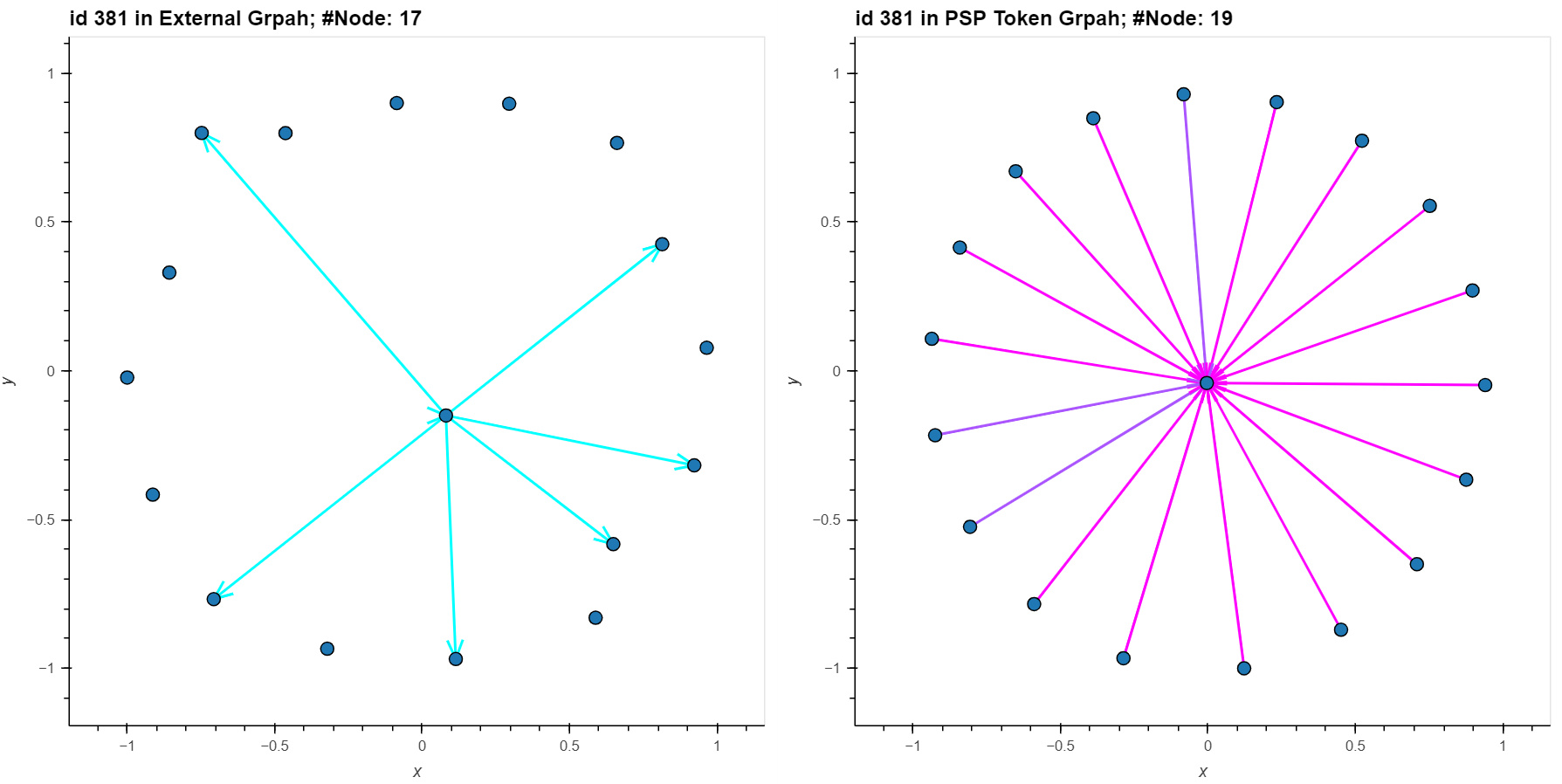}
    \caption{Clique ID-381}
    \label{fig:id-381}
    \Description{The left and right sides of Figure 13 represent the structure of the same clique in the external graph and the psp transaction graph, respectively. This is a sunflower structure with ID 381, which consists of 1 central point and 18 peripheral points. However, in the external graph, only the central node is connected to 6 other points. On the right side, all surrounding points are linked with center. More detailed description is in the text part of the paper.}
\end{figure}
This small group is a typical aggregation pattern with more than five participants judging from the PSP transaction graph on the right in Figure \ref{fig:id-381}. Only seventeen of the nineteen initial members were present in the external graph before 15$^{th}$ November, but they are not fully connected by edges. Based on the strong association of these addresses embodied in the PSP transaction graph on the right, it is reasonable to believe that other discrete nodes in the external graph are likely to be funded through CEXs, which means they can be funded without traces in our external transaction history or some other sponsoring node that strays from the airdrop reward list.

\subsubsection{The Blatant Hunter}
\label{sec:the_blatant_arbitrager}
\begin{figure}[htbp]
	\centering
    \includegraphics[width=\columnwidth]{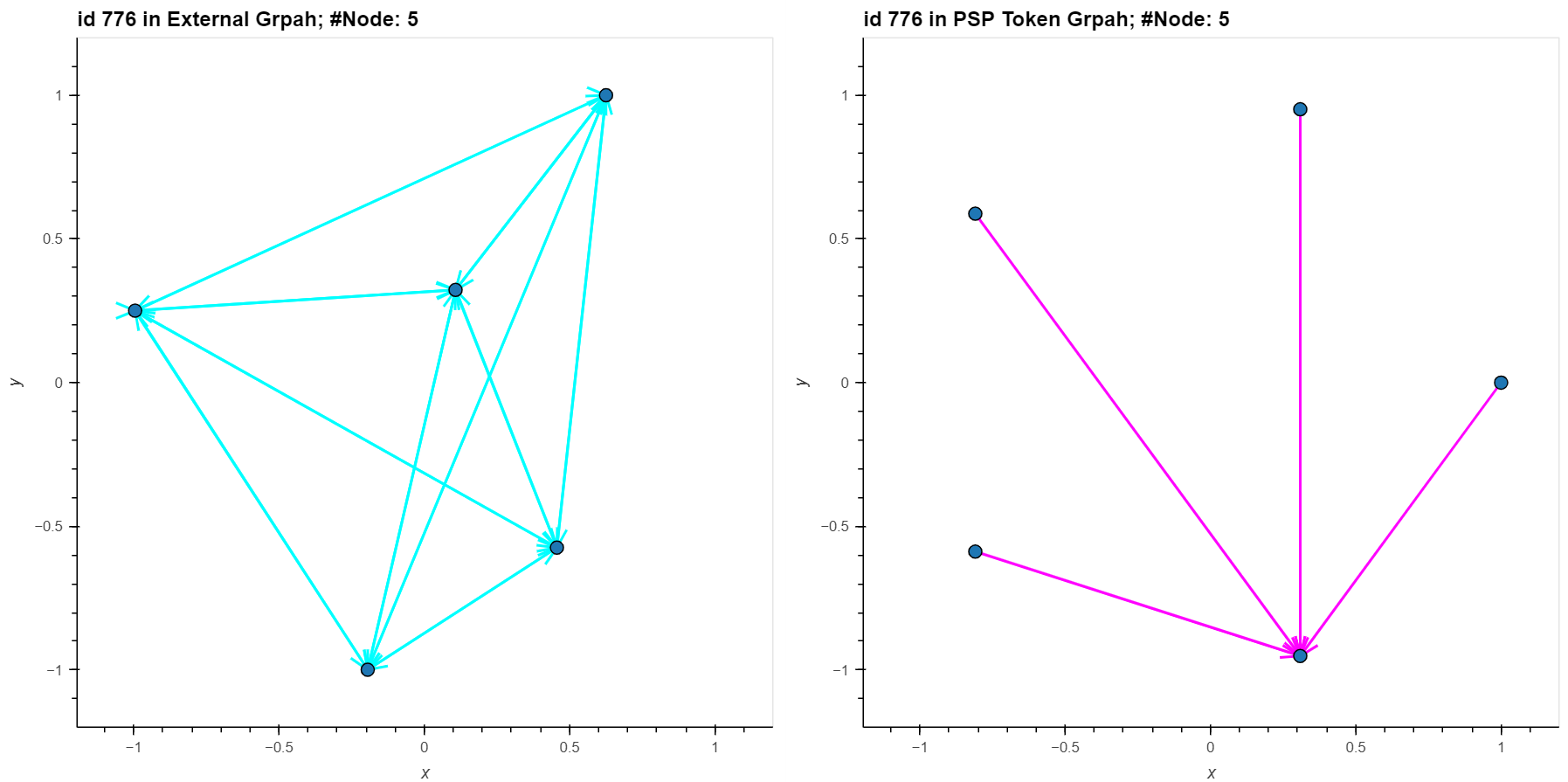}
    \caption{Clique ID-776}
    \label{fig:id-776}
    \Description{Figure 14 shows the Clique of ID-776, which consists of 5 points that are fully connected to each other in the external graph and unidirectionally connected to a central point in the psp transaction graph. More detailed description is in the text part of the paper.}
\end{figure}
Figure \ref{fig:id-776} exhibits part of the typical pattern of a large number of prevalent components that are heavily linked in the external graph and aggregate the resulting rewards to one initial member after the airdrop is issued. They are often composed of exactly five nodes, which is right under the detection conditions claimed by ParaSwap.

However, how should we think about this category of hunters who are on the edge of the rules? The contributions they make to qualify for airdrop meet the criteria of the community rules in monetary terms, but does this mean that the relationship between hunters and the community can be eased? Although, at this point, ParaSwap has not made its own screening code publicly available, we do not know what happened. But from a researcher's perspective, there is no doubt that arbitrage undermines the degree of decentralization that is the very core concept of blockchain platforms and Web3 communities.

\subsection{Discussion on Hunters and Network Analysis}
\label{sec:impact_of_arbitrage_cliques_on_airdrop_and_potential_solutions}
While there is much research now seeking to depict a vision of a harmonious, equal, accessible, and open Metaverse or Web3 community \cite{metaverse_1, metaverse_2}, we have to be reminded that we are dealing with a field that contains complex behavioral studies under anonymity, sophisticated economic models, and fierce mistrust as a foundation. Like the real-world problem of resource allocation and how government grants are distributed to stimulate economic recovery, this problem in the blockchain community may be a microcosm of the macroeconomic problem. Airdrop intends to increase user retention, distribute governance tokens to members who are genuinely engaged in the community, and decentralize DApps for stable long-term growth. However, the p2p cliques that we detected included more than half of the initial members, and the real situation is undoubtedly more than that, considering the prevalence of CEXs and mixing services \cite{mixing_services_www_2021}. The number of organizations or individuals operating behind these addresses will be far smaller than the number of users shown on the surface. With the presence of such a large percentage of addresses suspected of hunting airdrop, the stability of the airdrop currency issued would be detrimental or even disastrous.

The most immediate damage is monetary. As with stocks, a sell-off will bring down the price, but unlike stocks, free token trading on the blockchain does not have a mandatory circuit breaker mechanism. All holders of tokens will collectively face a price drop, at which point an avalanche-like chain reaction will send holders scrambling to sell in time to stop losses that can be enough to wipe out a token's value.

Systematically, the decentralization of DApps has also been undermined. It is worth noting that PSP Token also carries with it the responsibility of governance, meaning that in receiving rewards from airdrop, addresses simultaneously gain their respective degrees of voice and voting power, and holders are entitled to vote on decisions for the community. However, the presence of hunters occupies the rights that should be given to dozens of people into the hands of a small number of actual address holders, which is enough to cause the divergence or even the demise of a decentralized community, as the example mentioned of the Juno Network. When the decentralization of a DApp is null and void, and from the monetary aspect, it becomes unprofitable for its users, it is hard to imagine any other basis for its continuation. On a higher level, the instability caused by the challenges of decentralization will inevitably damage the credibility and growth potential of a DApp.


We believe that improvements can be made in terms of both screening methods and distribution policies. Although there is a nearly impenetrable and untraceable black box of CEXs, project owners should be able to track the network of more than second-degree relationships based on token transfers, thus excluding hunter cliques such as the "sponsor model" we find above. Or by comparing the behavior of addresses before and after the airdrop, using whether or not they are qualified as a label for the training set to obtain a classification algorithm for external graph features to filter at scale. In terms of the airdrop policy, we find that a tiered system of rewards will result in addresses with high rewards being more likely to engage in subsequent beneficial behaviors. More detailed and sophisticated tiered models may lead more participants to engage in positive activities like staking or providing liquidity.

Apart from the cold and emotionless code design and sophisticated model, we believe there is a more HCI way to the solution of bringing altruism into the Web3 community. We need to investigate how to reduce the generation of things like arbitrage, aggressive behavior, and inequality at the motivational level in a community with a high degree of anonymity. As pointed out by Elsden \emph{et al.} \cite{elsden2018making}, the most central issue appears to be some fundamental human challenges, including financialization \cite{financialization}, procedural trust, algorithmic governance, and front-end interactions. We suggest that in terms of Web3 community governance, the improvement can be conducted in two ways. The first is at the engineering level, where we argue that a figurative community enhances the sense of belonging of its members to a greater extent, which in return brings in more altruists. For example, a community of decentralized games can build stronger ties between players and content, and between players and players, than in a financial type DApp. Currently, no Metaverse-level platform can dominate and provide a space for digital people to exist in the true sense of the word. This means that a member in a loose community is nothing more than an unreachable piece of hexadecimal address to other members. Second, in terms of policy formulation, we argue that community leaders should always be wary of the negative effects of financialization. The blockchain platform is a huge and free market, and the impression it brings to the outside world has always been portrayed by the media as a shortcut to overnight riches, which in part has led to the current speculative culture in the entire field.

\section{Discussion and Future Vision}

Our research shows that, at this stage, decentralized communities are far from what they were designed to be. Airdrop hunters and speculators who might crowd out the space of community enthusiasts force other willing community members to give up their community-beneficial behavior. We conclude our result as challenges for future research with conceptual and methodological solutions to the current predicament.

\subsection{Expand the Ways of Interaction}
Studies on the influence of technologies and user experience (UX) design have suggested that a refined interaction can cause positive effects on user participation \cite{influencing_interaction, social_participation}. Off-the-shelf measurement frameworks have also long been available \cite{framework_HCI_participation}, but few of us bring the blockchain interaction down to the practical experiment. Although the blockchain network is claimed as Web3, from a user experience perspective, blockchain games, financial services, and other DApps are still in their primitive stage after so many years compared to the traditional Internet. Many studies have focused on designing user interactions for real-life Internet of Things applications rather than for virtual social network interactions \cite{promise_blockchain_interaction_design}.

We propose that in the future, with the rise of technologies such as VR/AR \cite{vrchat, ARPromoteSocialInteraction}, a pervasive and ambitious blockchain social network, such as the Metaverse, could be established to trigger users' motivation for engaging the community. However, it does not necessarily require cutting-edge technologies like VR/AR, a set of social layers of virtual identities on top of real ones while protecting anonymity can be applied and tested for its positive influence. For communities running on the particular infrastructure of blockchain, we currently know very little about the impact of interaction on these blockchain users and do not know if traditional approaches can motivate them or build strong connections or a sense of belonging between members and the community.

\subsection{Slow Down the Pace of Financialization}
As far as the decentralized community level is concerned, the idea of slowing down the pace of financialization is to separate the governance from tradable tokens. The financial attributes of blockchain cause a zero-sum game scenario that can significantly reduce the likelihood of cooperation among people. Giving the token both financial and governance features could be dangerous, as if allowing vote trading in real-life politics. The factual data results, as discussed in Section \ref{sec:datasets_and_experimental_setup}, show 86. 39\% of initial members sold out their PSP and thus left the community within six months. This phenomenon indicates that most of the initial members are attracted to the financial attributes of airdrop than the governance attributes. Considering the more than two million airdrop screening participants in the case of ParaSwap, financialization is an excellent incentive to attract users in the short term, however, in the long run, it cannot help retaining users and promote community engagement.

How do we design and evaluate a separate governance system? We consider that a new ERC token standard could be proposed to create a non-tradable, or worthless, voting ticket for participation in community governance, solving the problems in a blockchain way. However, separating the two would violate the design essence of property and ownership of digital assets. Another incentive mechanism can be applied in the form of rewards engaging the voting layer. Although what discussed above is just a brief idea, the key point is that the HCI community could play an important role in prototyping, envisioning, and evaluating such models.

\subsection{Improve the Model of Community Governance and Empowerment}
Ethereum offers any individual or organization a chance to establish their own circulable tokens with very relaxed restrictions on setting regulations and distribution models. It is the equivalent of a New York Stock Exchange with no vetting or restrictions, where anyone can conduct an initial public offering (IPO) to raise capital. Small and medium-sized teams with potential and strength have more opportunities to express their ambitions, but the majority left are frauds, bubbles, and Ponzi schemes \cite{ico_fraud, ico_scam}.

This hostile environment is one of the primary reasons preventing the development of stable decentralized communities. The efforts of researchers in this area should be instrumental in developing rules and policies, and the design of resource allocation by further understanding blockchain users and their roles. To give some simple, unexplored examples, we have difficulty understanding the behavior patterns of community builders and project initiators. Are there certain characteristics of their outreach on social media, such as Twitter? Do their conversations in discussion groups reflect potential fraud? What kind of white paper attracts investors the most? How does a committee of decentralized community work? Essentially, we try to show that exploring HCI in the blockchain is not limited to the user side but should also look broadly at other parties, such as rule makers and companies holding huge amounts of money, and design policy and resource allocation models that can coordinate the relationships between these parties.

Moreover, as outlined in the conclusion of Section \ref{sec:discussion_on_role_identification}, we proposed a multi-staged airdrop model that releases the second round reward based on the role identification of the initial round. Similarly, there is still significant potential in the  airdrop strategy and the process of empowering participants, which are vital elements in constructing decentralized communities.


\subsection{Understanding the Perception of Members in Decentralized Community}
Understanding users' behavior and values is crucial for incentivizing and building a community. If we can summarize their behavioral characteristics as well as the values and ideologies \cite{BitcoinIdeologicalTensions} driving behind them, perhaps the answer can be more easily obtained. On a cognitive level, the blockchain is still in its infancy and requires user education and a new way of thinking \cite{bitcoin_novice}. We lack evidence of the linkage between the user's level of professional knowledge and willingness to be a contributor. Will a sophisticated member be more profit-oriented? Or will a newbie dedicate more to the community? Although some blockchain-based social or instant messaging platforms have started to emerge in recent years, it is extremely difficult and time-consuming to investigate members of decentralized communities, especially those who are considered speculators and airdrop hunters behind the heavy curtain of anonymity.

As an example of our work, we could obtain more valuable information if a qualitative study on roles is conducted, but reaching a sufficient number of ParaSwap users to produce valid results is difficult given the background of the anonymous nature of blockchain. For role identification, although we have successfully labeled and categorized community members in terms of objective results, perhaps the categorization method would be more convincing if we recruited several users for each type of role for interviews and combined their motivations and behaviors for discussion. For clique detection, there is a high probability that airdrop hunters capable of conducting Sybil attack-like operations are people with considerable experience and background knowledge of blockchain technology, which could indicate that a profound understanding may not lead to an altruistic community member. However, this hypothesis can only remain at the conjecture stage until there is definitive evidence of the investigation. To use a more precise example, we want to know what factors lead to positive or negative behavior. For Diamond Holder, we want to understand their reasons for holding PSP for a long time and the motivating factors behind these reasons, such as the amount of digital assets they hold, their views on the future development of the project, etc. For Buyer, we can summarize their commonalities to deduce the reasons behind their buying behavior. In addition, when we find that interviewees have multiple addresses classified as different roles, qualitative study results can indicate users' habit of using multiple addresses.

In addition to surveys of certain specific users, a large-scale study on an open social platform or discussion group analysis would also be helpful. Jahani \emph{et al.} \cite{ScamCoins2018} conducted an exploratory study of the characteristics of online discussion around cryptocurrency forums. They concluded that projects with more information available and higher levels of technical innovation are associated with higher discussion quality through regression analysis. This kind of social computing research, which is not a stranger to us, would provide a wealth of topics worth exploring if it could be linked to the results generated from transaction records. For example, in a decentralized community with a high percentage of speculators and airdrop hunters, what kind of characteristics does the discourse on the forum have? Or conversely, we can explore whether a harmonious communication atmosphere is more likely to lead to an altruistic community environment.

\subsection{Future Challenge for Decentralized Community in Web3}
As aforementioned in Section \ref{sec:study_on_decentralized_communities}, unlike studies on community governance and moderation on Web 2.0 \cite{web2OnlineHarassment, web2RedditViolationScales}, most of the current blockchain community is more concerned with the development direction and resource allocation policy of DApps. However, with the rapid development of blockchain, some content-centric supporting infrastructures and corresponding streaming or social DApps such as Odysee\footnote{https://odysee.com/} are being developed \cite{web3LBRY}. In general, decentralized, online social media built with distributed storage as the back-end is not a new concept. They try to build a social network with no censorship, no single point of failure, and user autonomy that can support content verification through trust, and reputation \cite{web2DecentralizedADHocOSM}. Blockchain, in this situation, plays a fundamental role by providing a transparent, immutable underlying facility for information storage and retrieval \cite{web3SoChainDB}, proof as a service \cite{elsden2018making}, and incentives for providing content or authentication \cite{web3SocialRewardingMicrosopical}.

However, when DApps begin to evolve from such token exchange services into things that can be used by the average Internet user on a daily basis, away from the abstract virtual finance and cyberspace, how do these ahead-of-the-curve and idealistic products face real-world regulations? For example, Sağlam \emph{et al.} \cite{web3GDPR} pointed out that only 27.5\% of blockchain systems covered European Union's General Data Production Regulation (GDPR) through legal documents with privacy policies or terms and conditions. In addition to the internal aspect of privacy protection, the external threats of violating behavior, such as inflammatory speech, plagiarism, harassment, or even scams and crimes of cryptocurrency, still exist. Blockchain social network systems may introduce more new problems while keeping all the issues of the Web 2.0 community, which seems difficult to address, rely on loose community members without a robust executive body, or compromise the degree of decentralization. We propose that this specific moderation issue of users as shareholders on an equal footing with developers needs different ways of design and evaluation from the "Manage - Regulated" model of Web 2.0. But on the other hand, blockchain, as a new infrastructure, may technically give some existing Web 2.0 community moderation tools new possibilities and scope to perform. For example, for the governance of online speech \cite{web2TwitchShapingProAntiSocialBehavior}, the easy incentives facilitated by blockchain can support the establishment of positive role models or the tracing and discourage anti-social behavior with its public and transparent records. Mechanisms like governance tokens as rewards could also be used to select members from the community as stewards, giving them more power to further drive the community in a favorable direction.

But the challenge that we think is more worthy of attention is something more fundamental: the process of building community. We see our work not as being limited to a specific decentralized community or scenario, but as an exploration of the airdrop, as part of building the community and empowering its members. As the previous step to operating and governing a community, the growing stage of a decentralized community should be spent with the same research effort. While the vast majority of addresses exhibit arbitrage behavior, Web3 builders may need to rethink whether the users share their values and are ready for decentralization.

\section{Conclusion}
\label{sec:conclusion}

To sum up, in this paper, we use a detailed data-driven approach to understand altruistic and profit-oriented members in a decentralized community. We explore the rationality of airdrop as a kind of incentive mechanism and affirm the contribution of part of this allocation method: the differential tiers can be a potential method to encourage community-beneficial behaviors. We also perform a network analysis to identify the airdrop hunter's methods through multiple accounts. Our findings lead us to a discussion on the underlying causes of the current predicament and future direction of the decentralized community in Web3. We suggest that taxonomy of roles based on the public data could be a promising solution to support the empowerment process of the community. We propose that, beyond the technical difficulties, between the current decentralized community and the goals of the ideal Metaverse, there are still complex issues related to human factors, social organization, and financialized markets that require the participation of the HCI community. Moreover, with the development of Web3, we also raise potential governance issues regarding blockchain-based content platforms or social networks by analogy to Web 2.0. We hope that our work could play a role in advancing the understanding of this unique emerging technology and form of community, and further motivate researchers to push forwards the boundaries and expand the possibility of governance.

\begin{acks}
This work was supported by Shenzhen Science and Technology Program (Grant No. JCYJ20210324124205016).

\end{acks}



\balance{}

\bibliographystyle{ACM-Reference-Format}
\bibliography{sample-base}

\appendix

\end{document}